\documentclass[%
 reprint,
superscriptaddress,
 amsmath,amssymb,
 aps,
floatfix,
]{revtex4-2}
\usepackage[caption=false]{subfig}
\usepackage{graphicx}
\usepackage{dcolumn}
\usepackage{bm}
\usepackage{textcomp}
\usepackage{gensymb}
\usepackage{placeins}
\usepackage{hyperref}
\usepackage[utf8]{inputenc}
\usepackage{mathtools}

\newcommand{\subfigimg}[3][,]{
  \setbox1=\hbox{\includegraphics[#1]{#3}}
  \leavevmode\rlap{\usebox1}
  \rlap{\hspace*{0pt}\raisebox{\dimexpr\ht1-2\baselineskip}{#2}}
  \phantom{\usebox1}
}

\begin{document}

\title{Tune modulation effects in the High Luminosity Large Hadron Collider}

\author{S.~Kostoglou}
\email{sofia.kostoglou@cern.ch} 
\affiliation{
 CERN, Geneva 1211, Switzerland \\
}
\affiliation{
 National Technical University of Athens, Athens 15780, Greece\\
}
\author{H.~Bartosik}
\affiliation{
 CERN, Geneva 1211, Switzerland \\
}
\author{Y.~Papaphilippou}
\affiliation{
 CERN, Geneva 1211, Switzerland \\
}
 \author{G.~Sterbini}
\affiliation{
 CERN, Geneva 1211, Switzerland \\
}
\author{N.~Triantafyllou}
\affiliation{
 CERN, Geneva 1211, Switzerland \\
}
\affiliation{
 University of Liverpool, Liverpool L69 3BX, United Kingdom\\
}

\date{\today}

\begin{abstract}

Several transverse noise sources, such as power supply ripples, can potentially act as an important limiting mechanism for the luminosity production of the Large Hadron Collider (LHC) and its future High-Luminosity upgrade (HL-LHC). In the presence of non-linearities, depending on the spectral components of the power supply noise and the nature of the source, such a mechanism can increase the diffusion of the particles in the distribution through the excitation of sideband resonances in the vicinity of the resonances driven by the lattice non-linearities. For the HL-LHC, due to the reduction of the beam size in the Interaction Points (IP) of the high luminosity experiments (IP1 and 5), increased sensitivity to noise effects is anticipated for the quadrupoles of the inner triplets. The modulation that may arise from the power supply ripples will be combined with the tune modulation that intrinsically emerges from the coupling of the transverse and longitudinal plane for off-momentum particles through chromaticity. To this end, the aim of this paper is to study the impact of tune modulation effects on the transverse beam motion resulting from the interplay between quadrupolar power supply ripples and synchro-betatron coupling. A power supply noise threshold for acceptable performance is estimated with single-particle tracking simulations by investigating the impact of different modulation frequencies and amplitudes on the Dynamic Aperture. The excitation of sideband resonances due to the modulation is demonstrated with frequency maps and the higher sensitivity to specific modulation frequencies is explained. Finally, a power supply noise spectrum consisting of several tones is considered in the simulations to determine whether the presence of power supply ripples in the quadrupoles of the inner triplet will limit the luminosity production in the HL-LHC era.
\end{abstract}

\maketitle

\section{Introduction}

In order to optimize the performance of a high-energy particle collider such as the Large Hadron Collider (LHC) \cite{Bruning:782076}, a thorough understanding of all the phenomena that can degrade the luminosity is required. A major concern for the transverse single-particle beam dynamics is the presence of noise, a mechanism that can impact the long term stability of the circulating particles. Noise can arise from various sources such as fluctuations of the magnetic fields, ground motion and the transverse feedback system. It has been previously reported that such a mechanism can cause emittance growth \cite{lebedev, lebedev1,groundmotion, crabcavities}, particle losses \cite{Zimmermann:1993ny,SPSHERALHC} and eventually prove detrimental to the beam lifetime. 

From the plethora of noise sources that are present in the accelerator, this paper presents an approach to study the beam performance in the presence of a periodic perturbation in the strengths of the lattice magnets. Specifically, the analysis tools presented in this paper are employed to investigate the implications of power supply ripples in the quadrupoles located in the high \(\beta\)-function regions. Ripples in the power supply voltage are converted into current ripples, depending on the magnet's impedance, which eventually result in magnetic field perturbations through the magnet's transfer function (vacuum chamber, beam screen). As a result, these harmonic fluctuations induce a modulation in the normalized focusing strength of the quadrupoles. Depending on the frequency and the amplitude of the spectral components, a modulation of the betatron tune can arise with a modulation depth, i.e., the maximum tune variation from its unperturbed value, that is proportional to the \(\beta\)-function at the location of the perturbation. If present, this effect is combined with the tune modulation that intrinsically emerges for off-momentum particles from the coupling of the longitudinal and transverse plane through chromaticity \cite{synchrobetatron}. In the presence of non-linearities, tune modulation effects may lead to the excitation of sideband resonances, that, depending on the modulation frequency, can either overlap, leading to chaotic trajectories \cite{chirikov1960resonance, chirikov1971research}, or reach the tune footprint, thereby acting as an additional diffusion mechanism for the particles in the distribution \cite{HERA3}. In the latter case, the existence of such resonances critically limits the available space in frequency domain for an optimized, resonance-free working point. Subsequently, it is important to investigate whether the combination of the tune modulation induced by noise and synchro-betatron coupling will pose a limitation to the luminosity production of the future LHC upgrade, the High Luminosity LHC (HL-LHC) \cite{Apollinari:2284929}.

Tune modulation effects have been reported in the past from several hadron synchrotrons and colliders. In the Super Proton Synchrotron (SPS) at CERN, the large tune spread observed experimentally was associated with the presence of power supply ripples \cite{SPS1, SPS2}. During dedicated experiments, a tune modulation was introduced at various frequencies in the presence of significant non-linearities. A critical modulation depth threshold below \(\rm 10^{-3}\) was defined for acceptable performance. The combined effect of two modulation frequencies was experimentally proven to be more severe compared to a single frequency with an equivalent modulation depth. The presence of two modulation frequencies led to a more rapid lifetime decrease and a shift of the chaotic boundary towards lower amplitudes. The increase of the losses and emittance growth was correlated with the overlap of multiple resonances \cite{SPS3}. The studies conducted in the Relativistic Heavy Ion Collider (RHIC) at BNL investigated the combined effect of synchrotron motion and power supply ripples on the Dynamic Aperture (DA) \cite{RHIC4, RHIC3}. A linear correlation between the DA reduction and the synchrotron or gradient modulation depth was identified. By including in the simulations the voltage tones obtained from experimental observations, in combination with a modulation depth of \(\rm10^{-3}\) from synchro-betatron coupling, a DA reduction was observed for a variation of \(\rm10^{-5}\) in the quadrupolar current \cite{RHIC2}. In the Hadron-Electron Ring Accelerator (HERA) at DESY, a threshold of \(10^{-4}\) in the modulation depth was defined from analytical derivations in the presence of both beam-beam effects and a high-frequency tune modulation, a limit which was also verified experimentally by injecting external noise \cite{HERA1}. A dependence of the particles' diffusion on the modulation frequency was shown. For the first time, a tune ripple feedback was employed for the compensation of the power supply ripples by injecting an additional modulation with the same amplitude and an opposite phase \cite{HERA3, HERA2}. During these experiments, the use of such a compensation scheme proved to be beneficial for the beam lifetime.

The upgrade of the LHC to the HL-LHC aims to reach an integrated luminosity of 250 \(\rm fb^{-1}\) per year \cite{Apollinari:2284929}. To achieve this goal, the LHC will be operating with low beam emittances, high intensities, an unprecedented \(\beta ^*=15 \ \rm cm \) at the Interaction Points (IPs) of the high luminosity experiments (IP1 and 5), and thus strong beam-beam interactions. In addition to the incoherent effects, the suppression of the coherent beam motion requires the operation of the machine in a high chromatic and octupolar current regime for the mitigation of collective instabilities. This configuration results in significant non-linearities. Furthermore, a key component in the HL-LHC project is the upgrade of the Interaction Regions (IRs) around IP1 and 5. An important modification concerns the inner triplet layout, which provides the final focusing of the beam in the two high-luminosity experiments. Reducing the beam size at the two IPs results in the increase of the maximum \(\beta\)-functions at the quadrupoles of the inner triplet, rendering the beam more sensitive to magnetic field fluctuations at this location. Due to the combined effect of the significant non-linearities that arise from this configuration and the higher \(\beta\)-functions, a larger sensitivity to noise effects is anticipated. 

The unprecedented high-values of the maximum \(\beta \)-functions at these locations, in combination with the next triplet generation based on \(\rm Nb_{3}Sn\) technology that is currently being developed, justifies the need to perform a complete analysis on the beam performance implications of a modulation in the quadrupole gradients. Therefore, our investigation focuses on the four quadrupoles, namely Q1, Q3 (MQXFA magnets) and Q2a, Q2b (MQXFB magnets), circuits which are electrically powered in series. Previous studies for the HL-LHC have revealed a dependence of the DA reduction on the modulation frequency and specifically, a more significant effect to noise at 300~Hz and 600~Hz \cite{HL_LHC_rip}. In the context of this investigation, the aim of the present paper is to explain the sensitivity to power supply ripples at particular frequencies and to quantify their impact with 5D and 6D single-particle tracking simulations, including beam-beam effects, in terms of tune diffusion, intensity evolution and DA.

This paper is organized as follows. In Section~\ref{henon} a simplified formalism with a linear decoupled rotation, a modulated quadrupole and non-linear elements is employed. Starting from the characterization of the single-particle spectrum and extending to a distribution of particles, the impact of the modulation in configuration space and frequency domain is reported. These observations are then validated with tracking simulations in the HL-LHC lattice, including power supply ripples in the inner triplets of the high luminosity experiments and/or synchro-betatron coupling in Section~\ref{sec:HL-LHC}. The expected impact on the tune diffusion and intensity evolution is evaluated to determine whether power supply ripples in the triplet will be an important concern for the beam performance in the future operation.

\section{Modulated simplified map}
\label{henon}

\begin{figure}
\includegraphics[width = 0.9\columnwidth]{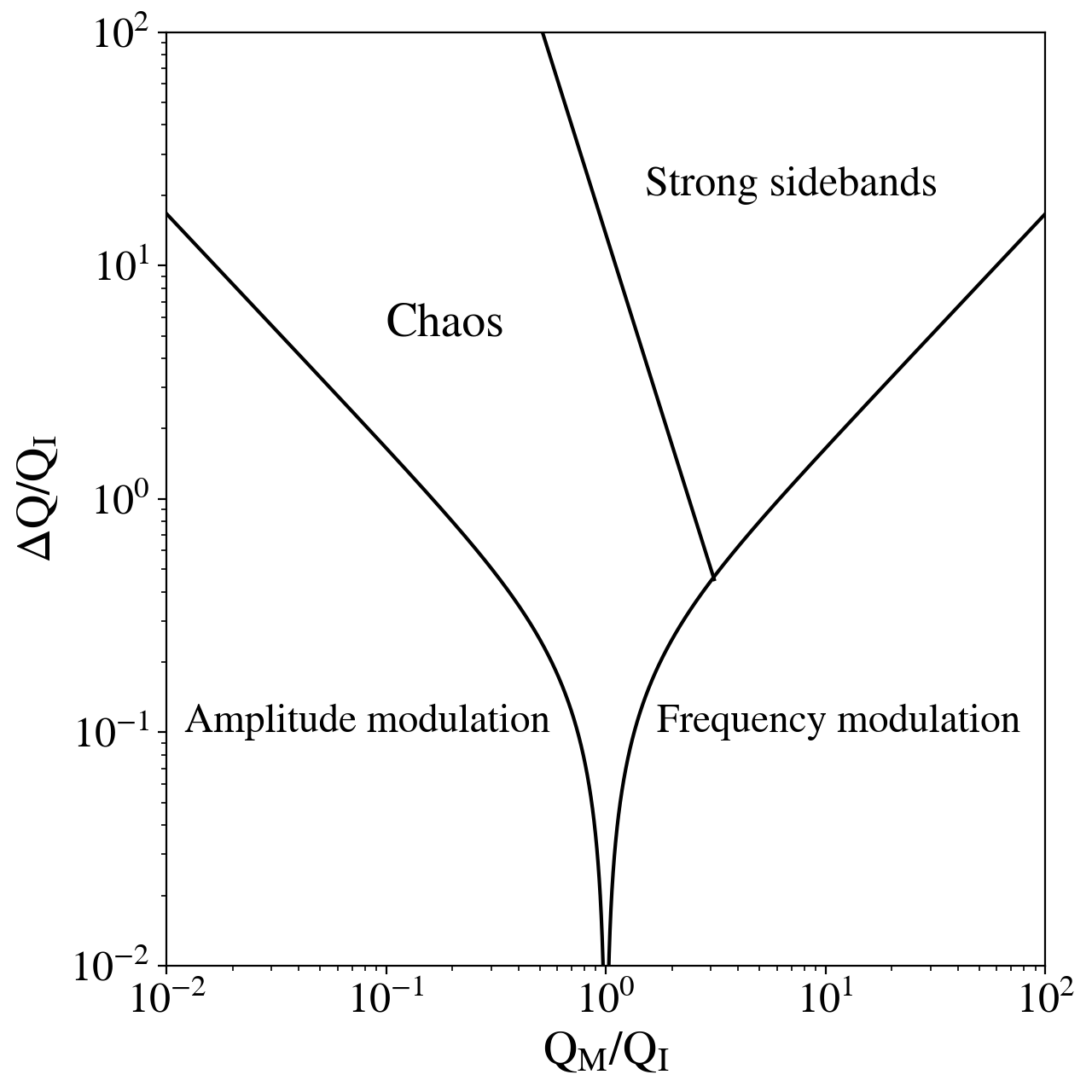}
\caption{\label{fig:tune_mod_parameter_space} The four regimes in the tune modulation parameter space for a modulation tune $Q_m$ and depth $\Delta Q$ with an island tune $Q_I$ \cite{Satogata:1993kv, satogata1991beta, Zimmermann:1993ny}.}
\end{figure}

The effect of a tune modulation in a one-dimensional non-linear Hamiltonian system has been extensively studied in the past \cite{Satogata:1993kv, satogata1991beta, Zimmermann:1993ny}. These studies have shown that, depending on the modulation tune, \(Q_m\), and modulation depth, \(\Delta Q\), the impact on the transverse motion varies. Through an analytical formalism, in the presence of a single resonance with an island tune $Q_I$ and a tune modulation, four regimes of interest were identified in the tune modulation \((Q_m, \Delta Q)\) parameter space \cite{Satogata:1993kv, satogata1991beta, Zimmermann:1993ny}. As an example, Fig.~\ref{fig:tune_mod_parameter_space} illustrates the four regimes in the tune modulation parameter space for the sixth-order resonance. First, a modulation at low amplitude and low frequency causes a time variation of the particles' actions, without affecting their stability (\textit{Amplitude modulation} regime). Second, increasing the frequency leads to the appearance of sideband islands around the resonance, in a distance equal to the modulation tune. For small modulation depths, the islands occupy a limited portion of the phase space (\textit{Frequency modulation}). As the modulation depth increases, high-order sideband islands become significant (\textit{Strong sidebands}). The combination of a strong modulation in terms of amplitude and a frequency below a critical value, results in the overlap of the resonance islands, hence leading to chaotic motion (\textit{Chaos}). 

The existence of islands in the phase space originating from the strong non-linearities of the lattice was experimentally demonstrated through observations of persistent signals in the beam position measurements \cite{Tevatron1}. Then, by injecting noise that resulted in a tune modulation, the particles were driven outside of the resonance islands. Based on the beam response, the existence of the three out of four regimes in the tune modulation parameter space was experimentally verified \cite{Tevatron2}. In this context, the present paper investigates the effect of power supply ripples in the voltage-controlled regime, which extends beyond a critical value of a few Hz \cite{Martino}. Subsequently, the regimes of interest in the aforementioned parameter space are the \textit{Frequency modulation, Strong sideband} and \textit{Chaos}, where emittance growth and particle losses occur. The \textit{Amplitude modulation} regime is not discussed in the present paper as it mainly refers to low-frequency noise. 

This section focuses on the description of the particles' motion under the influence of a modulated quadrupole using a simplified formalism based on a linear decoupled rotation and non-linear elements. As a starting point, a 2D model is employed to illustrate the modulation in frequency domain in the absence of non-linearities. The impact of the three regimes of interest in the tune modulation parameter space is then summarized by accounting for non-linearities. The transfer map is then extended to 4D to study the instantaneous variation of the tune footprint and the sideband resonances with frequency maps.

\subsection{Tune modulation in 2D}
\label{tune_mod_2D}
\subsubsection{Instantaneous tune}
As a first step, a single particle with a zero initial transverse momentum is tracked in a lattice that consists of a linear rotation and a modulated quadrupole (without including non-linearities). A tune modulation is induced by varying the quadrupolar strength turn-by-turn with a sinusoidal function at a single frequency, rather than directly modulating the tune in the rotation matrix. The variation of the tune, referred to as the instantaneous tune, as a function of the turn number $n$ is represented by:
\begin{equation}
    Q_{inst}(n) = Q_0 + \Delta Q \cdot \cos(2\pi Q_m n),
\label{eq:inst}
\end{equation}
where \(Q_0\) is the unperturbed tune in the absence of the modulation, $\Delta Q$ and $Q_m$ is the modulation depth and modulation tune, respectively. The rotation is divided into eight segments with equal values of the betatron phase advance and the position measurements are retrieved after each segment. This method provides a uniform sampling for the analysis in frequency domain. In this way, the sampling rate increases by a factor of eight and an accurate tune determination with a limited number of turns is achieved using the implementation of the Numerical Analysis of Fundamental Frequencies (NAFF) algorithm \cite{NAFF, NAFF1, PhysRevAccelBeams.22.071002}. An approximation of the instantaneous tune is computed with a sliding window of 30 turns in steps of one turn. Such a window length was found to be a good trade-off between time and frequency resolution for the tune determination.

\begin{figure}
\subfloat{\subfigimg[width=0.98\columnwidth]{\textbf{a)}}{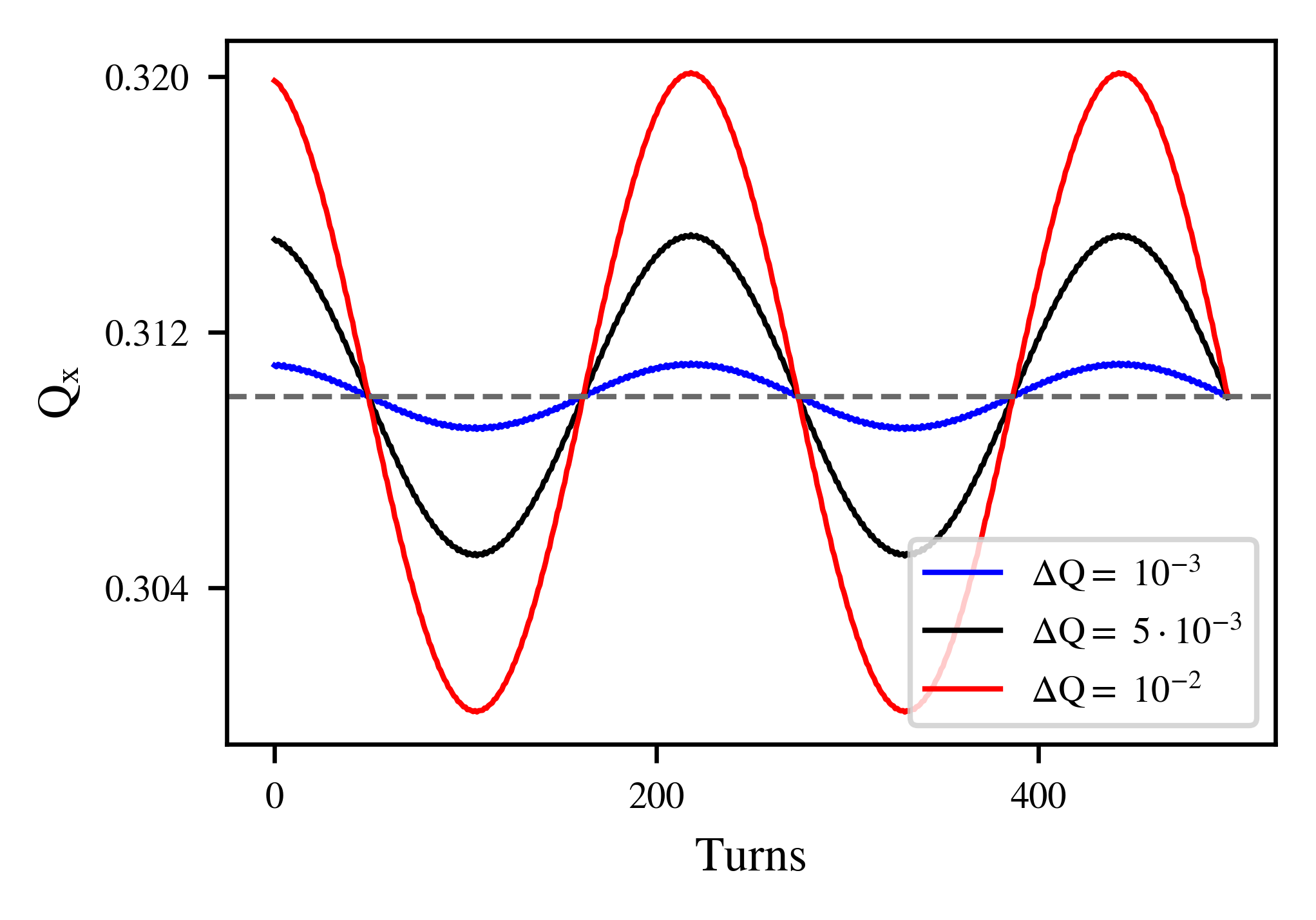} \label{subfig:mod1}} \\
\subfloat{\subfigimg[width=0.98\columnwidth]{\textbf{b)}}{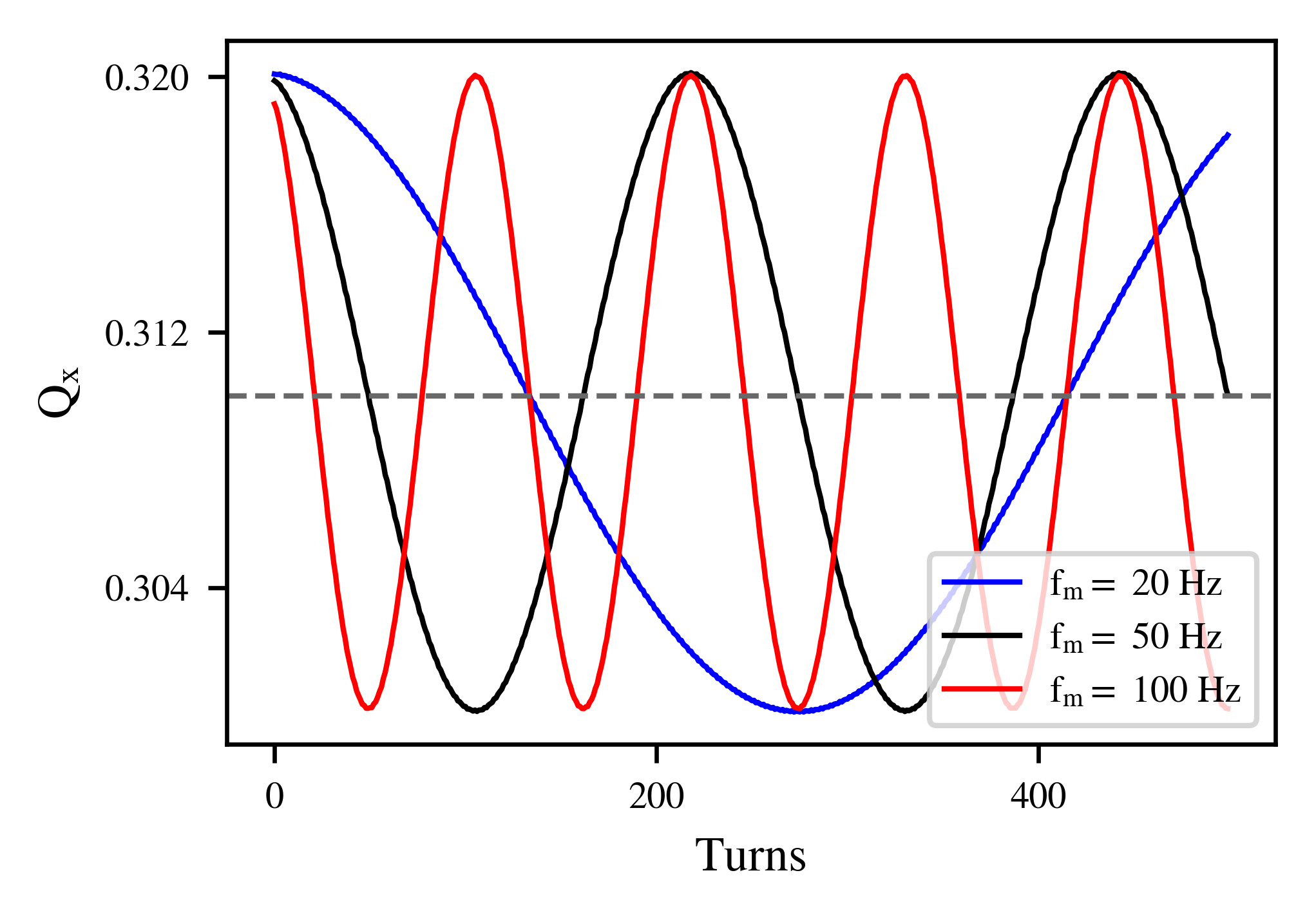} \label{subfig:mod2}}
\caption{\label{fig:instanteneous} (a) An approximation of the instantaneous tune in the presence of a modulated quadrupole for \(f_m= \rm 50 \ Hz\) and \(\Delta Q=\rm 10^{-3}\) (blue), \(\Delta Q=\rm 5 \cdot 10^{-3}\) (black) and \(\Delta Q=\rm 10^{-2}\) (red). (b) An approximation of the instantaneous tune for \( \Delta Q=\rm 10^{-2}\) and \(f_m=\rm 20 \ Hz\) (blue), \(f_m= \rm 50 \ Hz\) (black) and \(f_m=\rm 100 \ Hz\) (red). The horizontal gray line denotes the unperturbed betatron tune.}
\end{figure}

The two parameters of interest are the modulation depth $\Delta Q$ and modulation tune $Q_m$, the ratio of which defines the modulation index $\beta_m$. Figure~\ref{fig:instanteneous} demonstrates the evolution of the instantaneous tune for an increasing value of the modulation depth (Fig.~\ref{subfig:mod1}), considering a constant frequency, and for several values of the modulation frequency with a constant modulation depth (Fig.~\ref{subfig:mod2}), assuming the LHC revolution frequency (\(f_{\rm rev} = \rm 11.245~kHz\)). The gray dashed line represents the unperturbed tune in the absence of a tune modulation. In particular, in the case of a tune modulation induced by noise in a quadrupole, the modulation depth is proportional to the \(\beta\)-function at the location of the perturbation and the maximum variation of the quadrupole normalized strength. Similarly, in the presence of synchro-betatron coupling, the modulation index depends on the chromaticity, the momentum deviation of the particle and, in this case, the synchrotron tune is the modulation frequency \cite{Suzuki:201687}. It must be noted however, that for a matched bunch, as all the particles in the distribution experience the same quadrupolar modulation, this effect can be measured. On the contrary, due to the different momentum deviations of the particles in the distribution, synchro-betatron coupling is more difficult to measure experimentally.

\subsubsection{Average single-particle spectrum}
\begin{figure*}
\subfloat{\subfigimg[width=\columnwidth]{\textbf{a)}}{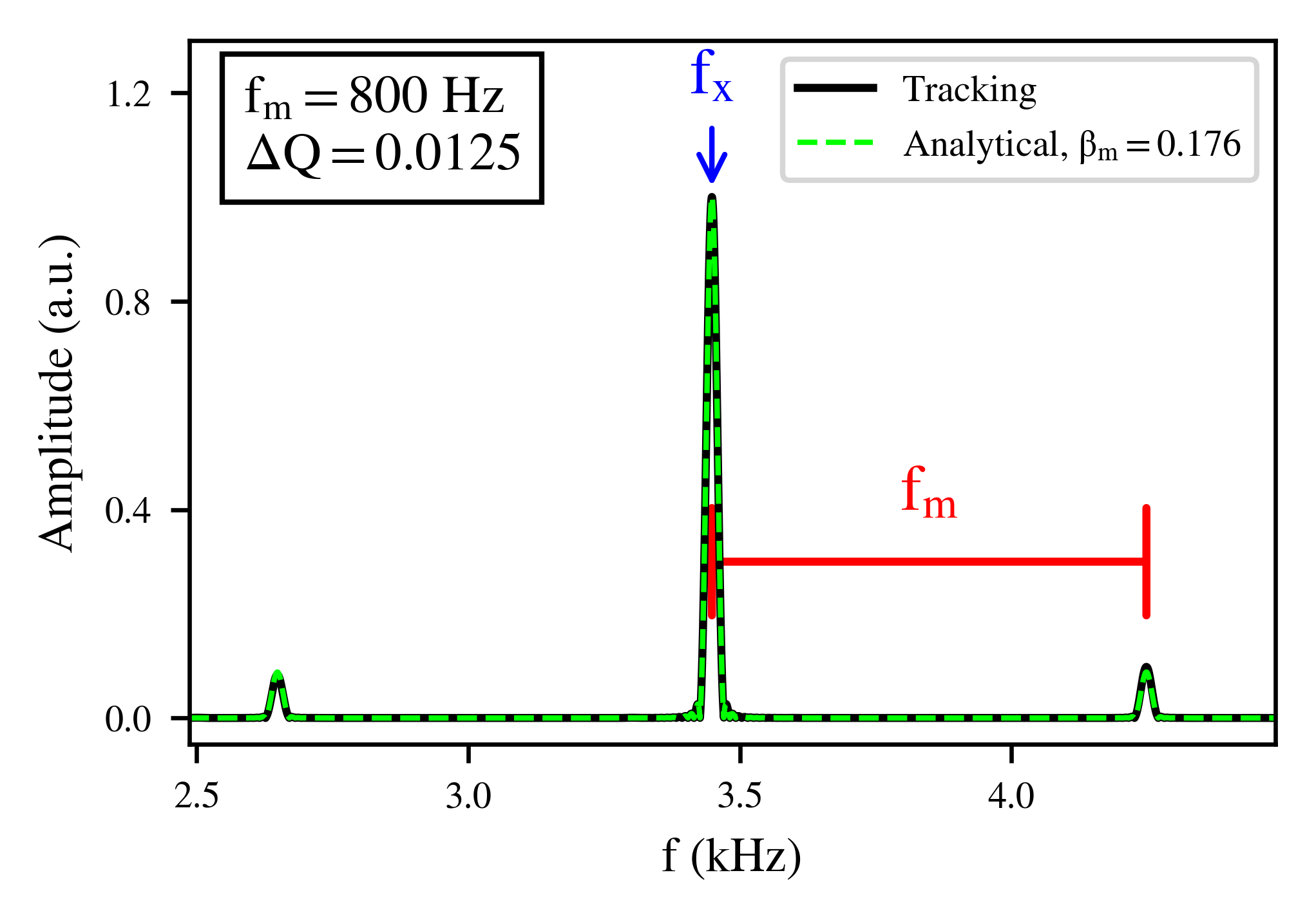} \label{subfig:fi1}}
\subfloat{\subfigimg[width=\columnwidth]{\textbf{b)}}{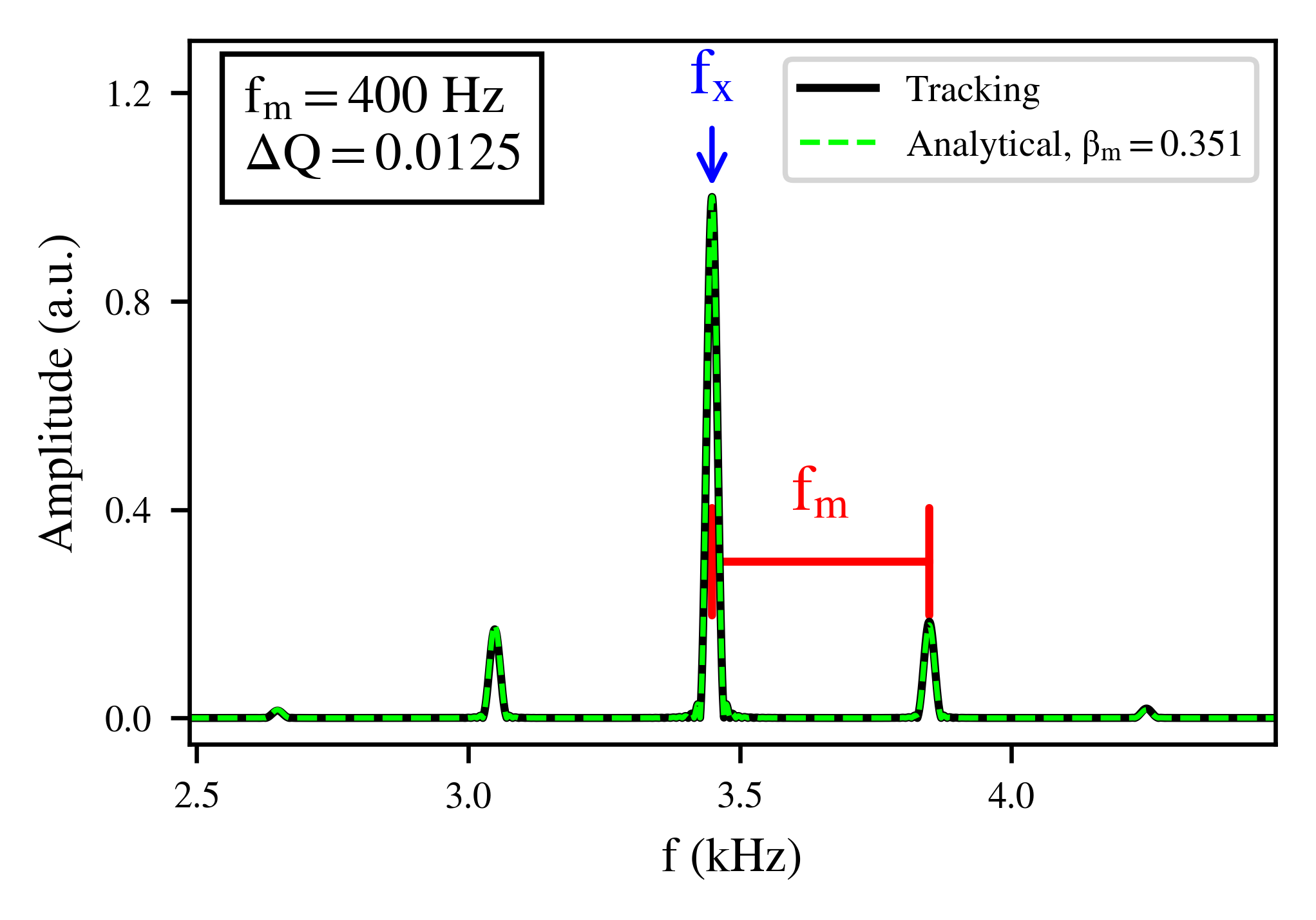} \label{subfig:fi2}} \\ 
\subfloat{\subfigimg[width=\columnwidth]{\textbf{c)}}{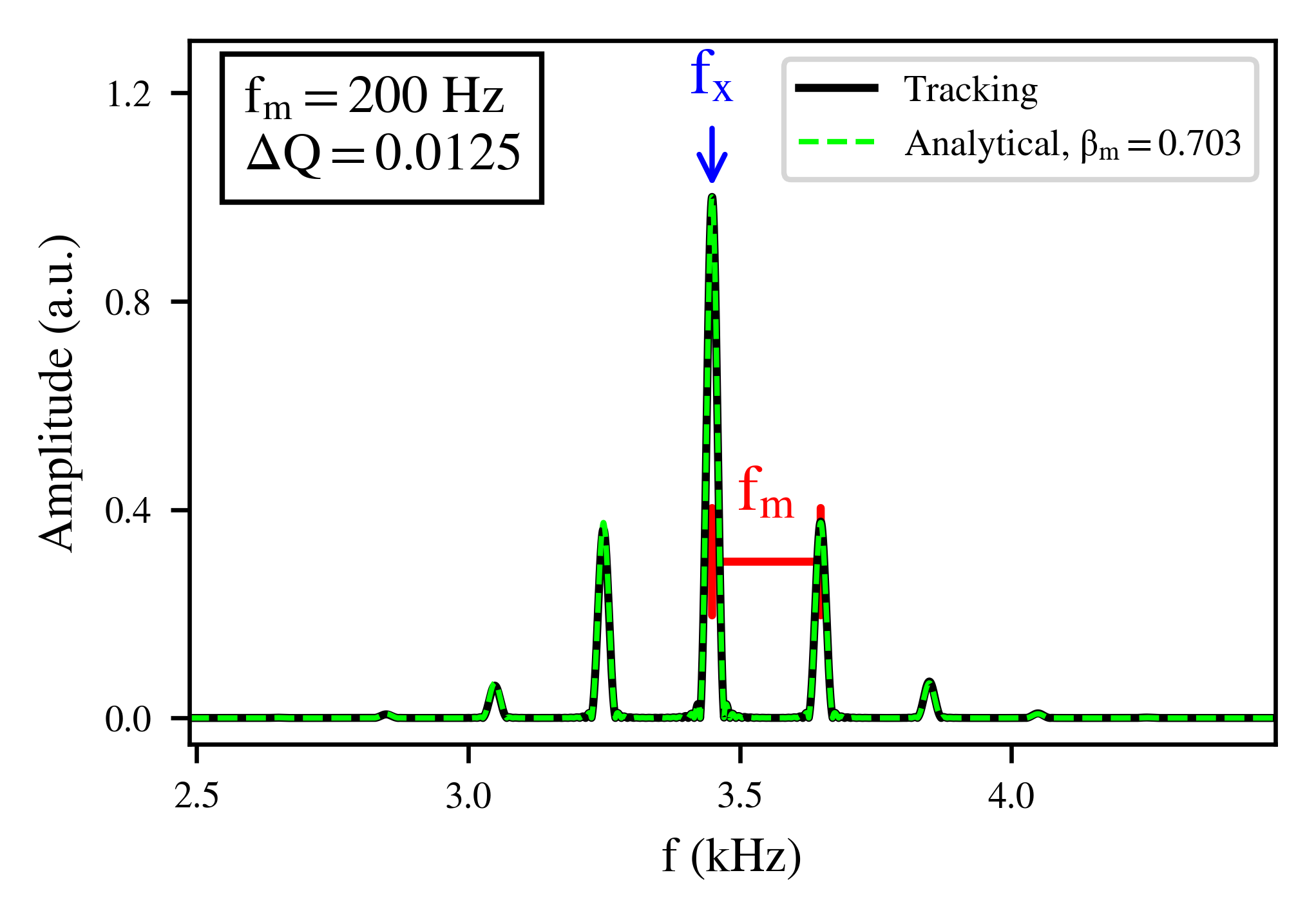} \label{subfig:fi3}}
\subfloat{\subfigimg[width=\columnwidth]{\textbf{d)}}{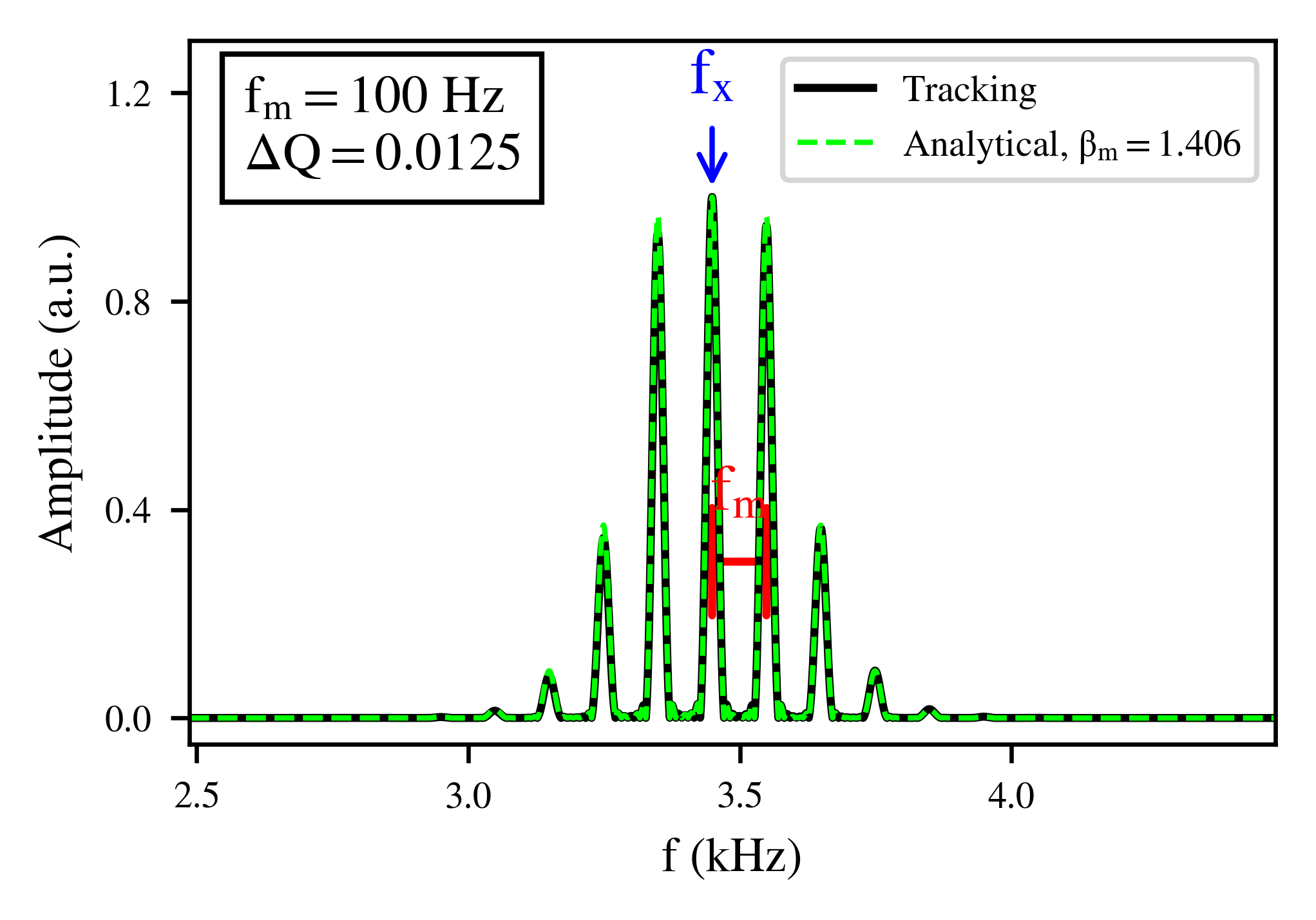} \label{subfig:fi4}}  \\
\subfloat{\subfigimg[width=\columnwidth]{\textbf{e)}}{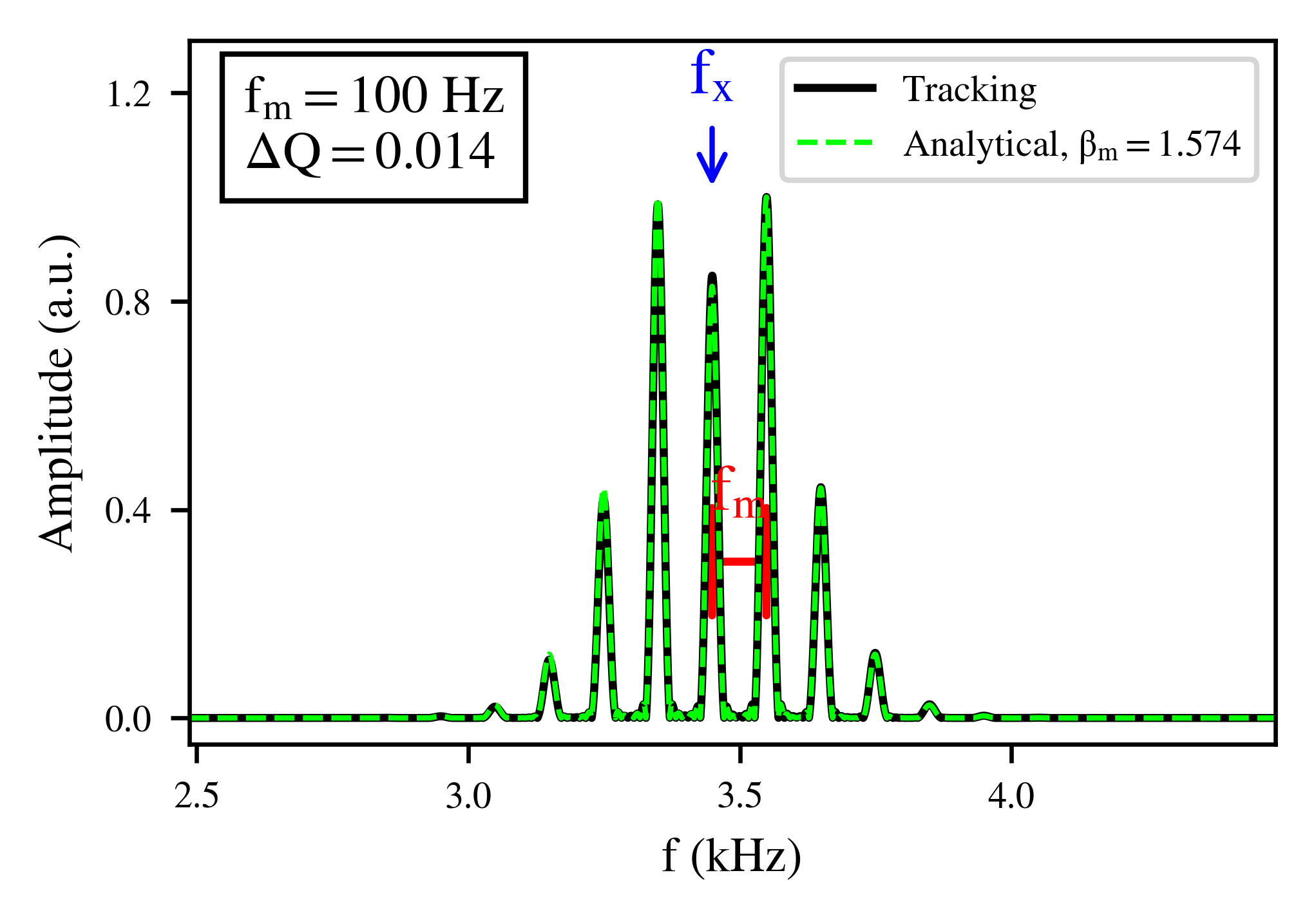} \label{subfig:fi5}}
\subfloat{\subfigimg[width=\columnwidth]{\textbf{f)}}{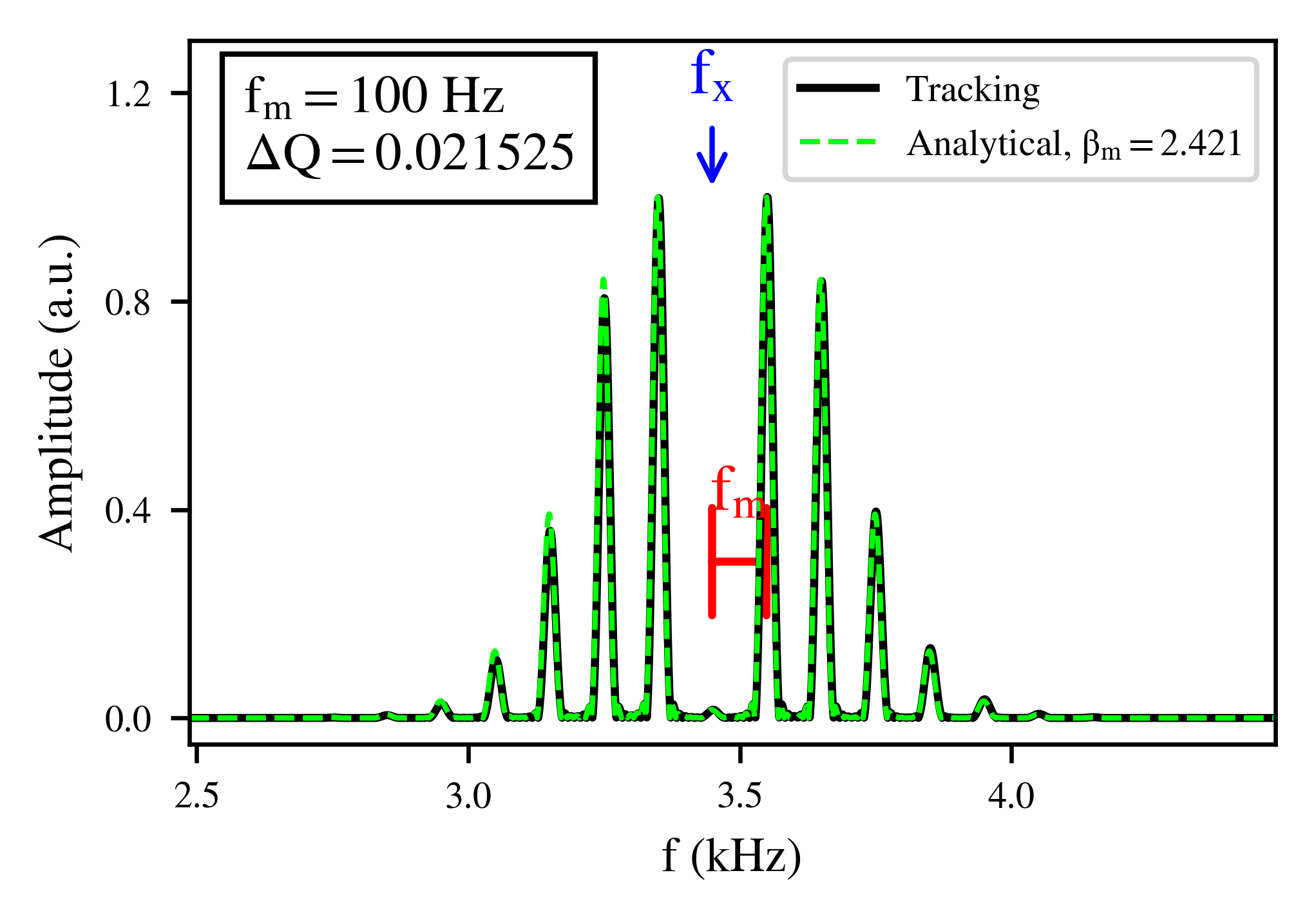} \label{subfig:fi6}} 
\caption{\label{fig:bessel} The normalised single-particle spectrum in the presence of a tune modulation at (a) 800~Hz, (b) 400~Hz, (c) 200~Hz and (d) 100~Hz and \(\Delta Q = \rm 1.25\cdot 10^{-2}\) and with (e) \(\Delta Q = \rm 1.4\cdot 10^{-2}\), (f) \(\Delta Q =\rm 2.2\cdot 10^{-2}\) at 100~Hz. The green curve indicates the representation of the particle's motion with the sum of Dirac functions weighted with the Bessel functions of the first kind for the various modulation indexes. The betatron frequency (blue) and its distance from the first positive sideband (red) are also depicted.}
\end{figure*}

To illustrate the importance of the modulation index and its impact on the particle's motion, the tracking is extended to \(\rm 10^4\) turns. During this time span, the variation of the tune depicted in Fig.~\ref{fig:instanteneous} is averaged over several modulation periods. As in every frequency modulated signal, considering the betatron motion as the carrier and the power supply ripples as the modulator, the particle's trajectory is represented in time domain by a sum of co-sinusoidal signals, weighted by the Bessel functions of the first kind \cite{roder1931amplitude}. By integrating Eq.~\eqref{eq:inst}, the frequency-domain representation of the betatron motion, after normalizing by the maximum amplitude, is:
\begin{equation}
\begin{split}
  |\tilde{X}(Q)| = \sum_{k=-\infty}^{\infty}J_k(\beta_m)[\delta(Q-Q_0-k\cdot Q_m) + \\
  \delta(Q+Q_0+k\cdot Q_m)],   
  \end{split}
  \label{eq:bessel}
\end{equation}
where \(\beta_m=\frac{\Delta Q}{Q_m}\) is the modulation index, \(J_k\) are the Bessel functions of the first kind and $\delta$ is the Dirac function.
Consequently, the spectrum consists of infinite harmonics of the modulation frequency around the betatron frequency with a relative amplitude that depends on the modulation index. In particular, the \(k\)-order Bessel function for a modulation index \(\beta_m\) determines the amplitude of the $k$-order sideband, with \(k=0\) representing the betatron tune. Figure~\ref{fig:bessel} presents the Fourier spectrum as computed with the NAFF algorithm (black) and analytically from Eq.~\eqref{eq:bessel} (green). In the first four cases, the modulation depth is constant, while the frequency reduces from 800~Hz (Fig.~\ref{subfig:fi1}) to 400~Hz (Fig.~\ref{subfig:fi2}), 200~Hz (Fig.~\ref{subfig:fi3}) and 100~Hz (Fig.~\ref{subfig:fi4}). In each case, sidebands around the betatron tune (blue) are present in the spectrum at a distance equal to multiples of the modulation frequency. The distance of the first positive sideband from the betatron frequency is also depicted (red). Based on the fact that the modulation index is inversely proportional to the modulation frequency, decreasing the modulation frequency for a constant modulation depth leads to an increase of the modulation index. Subsequently, higher-order sidebands with larger amplitudes are visible in the spectrum for lower frequencies. 

Then, for a constant modulation frequency at 100~Hz, the modulation depth is increased (Fig.~\ref{subfig:fi4} to \ref{subfig:fi6}). Reviewing the spectra shows that for \(\beta_m\geq\)1.5, due to the strong modulation, the amplitude of the sidebands exceeds the one of the betatron tune (Fig.~\ref{subfig:fi5}) and for specific values of the modulation index the amplitude of the betatron peak can also be suppressed (Fig.~\ref{subfig:fi6}). In such cases, using a peak-detection algorithm that sorts the frequencies by decreasing amplitude, such as NAFF, will return the frequency determination of the sideband peak rather than the one of the betatron tune. To overcome this problem, in the presence of a strong modulation, the algorithm's frequency range of search should be limited in the vicinity of the betatron tune. This is not always possible as for small modulation frequencies, the limited resolution does not allow to disentangle the sideband from the betatron peak. For instance, in the LHC, the modulation induced by the synchrotron motion ($f_m$ $\approx$20~Hz) to a particle which is placed in the limit of the bucket height ($\Delta p/p=27 \cdot 10^{-5}$) and in the presence of a high chromaticity ($Q'=15$) leads to a modulation index which exceeds the critical value of 1.5 ($\beta_m=2.25$). For the purpose of the study presented in this paper, the tracking is, in this case, performed either in 4D or 5D, i.e., neglecting the impact of the synchrotron motion or in 6D with a lower chromaticity and thus, a lower modulation depth.

\subsubsection{Tune modulation and non-linearities}
\begin{figure*}
\subfloat{\subfigimg[width=0.3\textwidth]{\textbf{a)}}{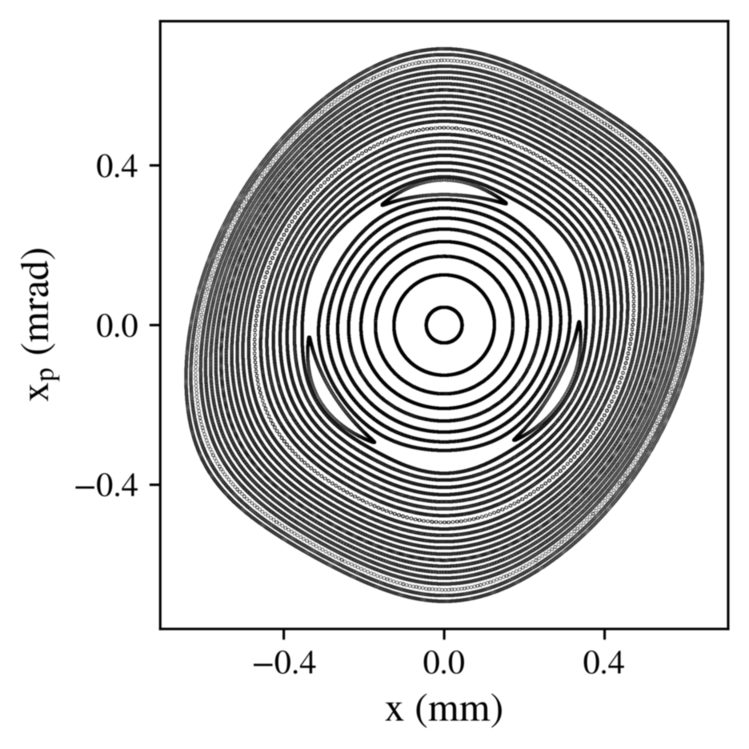} \label{subfig:phase_space_0}} 
\subfloat{\subfigimg[width=0.3\textwidth]{\textbf{b)}}{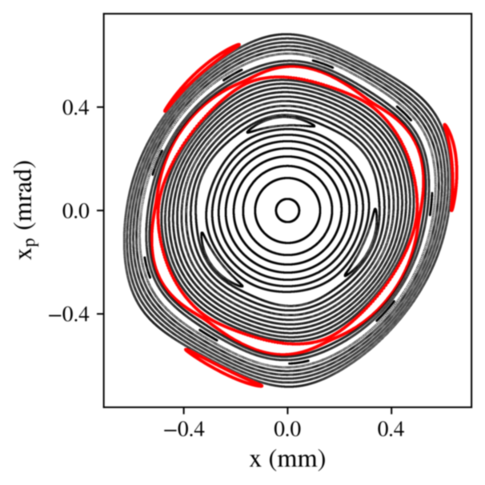} \label{subfig:phase_space_1}}  
\subfloat{\subfigimg[width=0.3\textwidth]{\textbf{c)}}{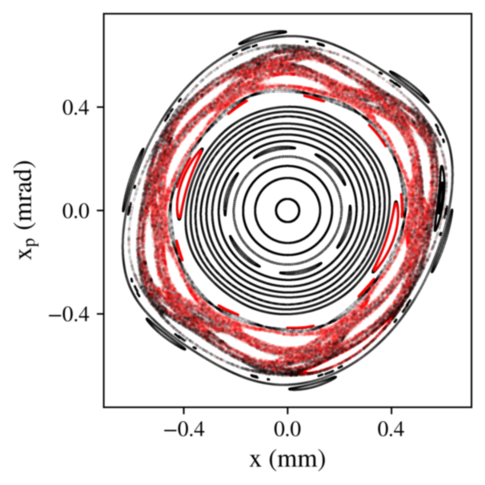} \label{subfig:phase_space_2}}
\caption{\label{fig:phase_space} The phase space in the presence of non-linearities (a) without a tune modulation, (b) with a tune modulation in the \textit{Frequency modulation} regime and (c) in the \textit{Strong sidebands}/\textit{Chaos} regime.}
\end{figure*}

Non-linear fields are an important aspect of the accelerator's lattice, introduced for the correction of the chromatic aberrations and the action-dependent detuning of the particles. Of particular interest in this study is the interplay between non-linearities and the tune modulation. To illustrate this effect, first, an octupolar element is included in the map and the working point is moved in the vicinity of a horizontal sixth-order resonance. A distribution of particles with linearly increasing horizontal actions is tracked in the simplified lattice. First, Fig.~\ref{subfig:phase_space_0} shows the turn-by-turn phase space, which reveals that several particles are trapped in the resonance islands. Second, a modulated quadrupole is included, with a combination of parameters \((Q_m, \Delta Q)\) that correspond to the \textit{Frequency modulation} regime. For the Poincar\'e section, the stroboscopic technique is used, where the horizontal coordinates are plotted once per modulation period. The phase space depicted in Fig.~\ref{subfig:phase_space_1} reveals the existence of sideband islands (red) in the vicinity of the ones presented in Fig.~\ref{subfig:phase_space_0}. Finally, the parameters of the modulation in Fig.~\ref{subfig:phase_space_2} correspond to the limit of the \textit{Strong sideband} and \textit{Chaos} regime. In this case, the simultaneous reduction of the modulation frequency and increase of the modulation depth results in overlapping higher-order sidebands (red), thus creating layers of chaotic motion. 

\begin{figure}
\subfloat{\subfigimg[width=0.98\columnwidth]{\textbf{a)}}{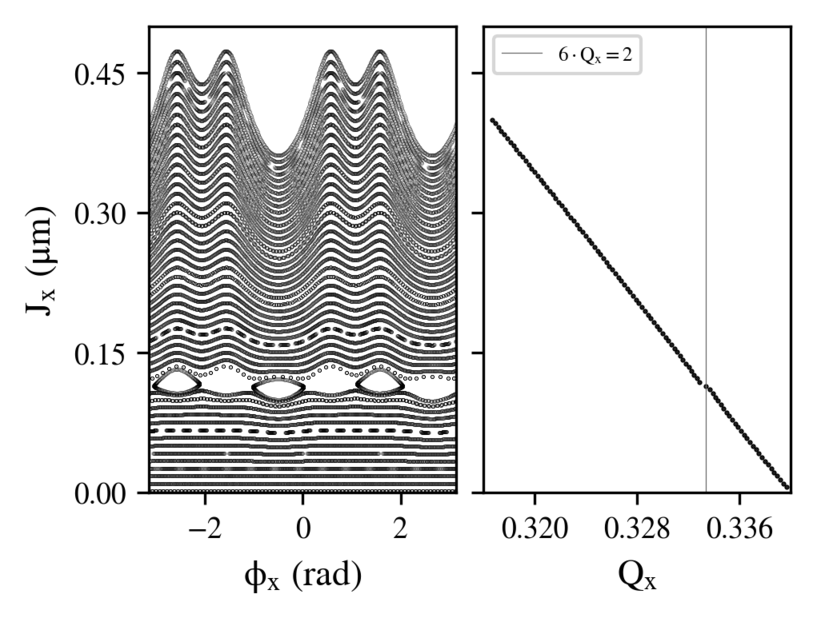} \label{subfig:action_angle_0}} \\
\subfloat{\subfigimg[width=0.98\columnwidth]{\textbf{b)}}{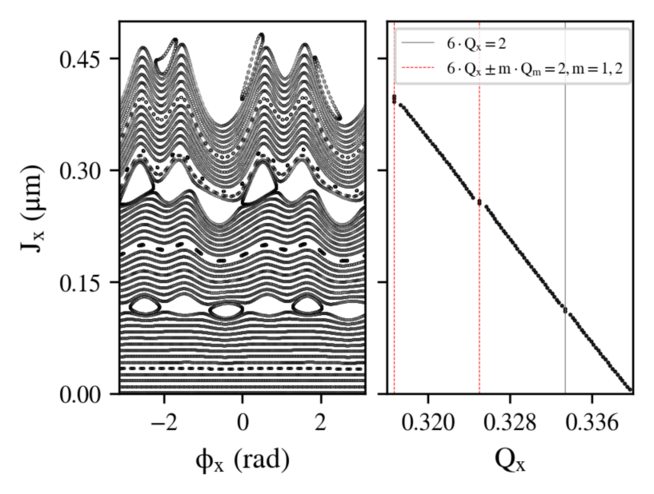} \label{subfig:action_angle_1}} \\ 
\subfloat{\subfigimg[width=0.98\columnwidth]{\textbf{c)}}{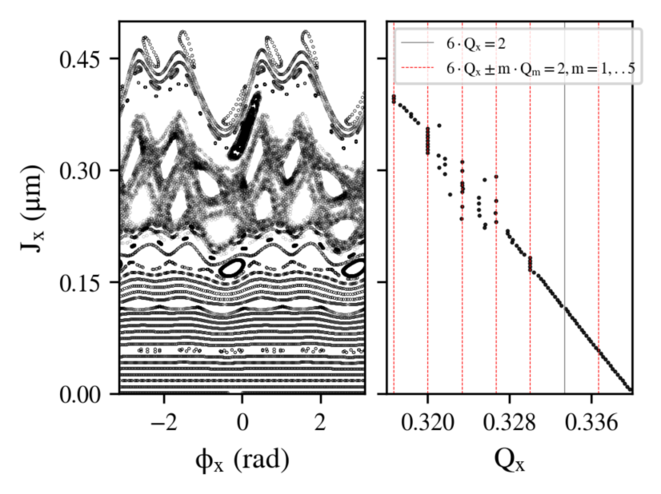} \label{subfig:action_angle_2}}
\caption{\label{fig:action_angle} Action-angle variables (left panel) in the presence of non-linearities (a) without a tune modulation, with a tune modulation (b) in the \textit{Frequency modulation} regime and (c) in the \textit{Strong-sideband}/\textit{Chaos} regime, next to the tune of each particle as a function of its action (right panel). The vertical lines represent the nominal (gray) and sideband (red) resonances.}
\end{figure}

Similarly, the action-angle variables are computed for each of the aforementioned cases. Furthermore, the tune of each particle is computed with the NAFF algorithm in order to illustrate the position of the sideband islands in frequency domain. Figure~\ref{fig:action_angle} depicts the action-angle variables (left panel) and the tune as a function of the action (right panel). In the absence of a tune modulation (Fig.~\ref{subfig:action_angle_0}), the islands correspond to the trapping of the particles in the sixth-order resonance (vertical gray line). In the \textit{Frequency modulation} regime (Fig.~\ref{subfig:action_angle_1}), the tune determination indicates that the additional resonances are the first and second-order sideband of the modulation frequency (red vertical lines). The additional resonances are located in a distance equal to multiples of the modulation frequency. In the last case (Fig.~\ref{subfig:action_angle_2}), the particles are trapped up to the fifth-order sideband of the modulation frequency and the chaotic motion of the particles is also visible in their tune determination.

\subsection{Tune modulation in 4D}
\subsubsection{Instantaneous tune footprint}
\begin{figure}
\includegraphics[width = \columnwidth]{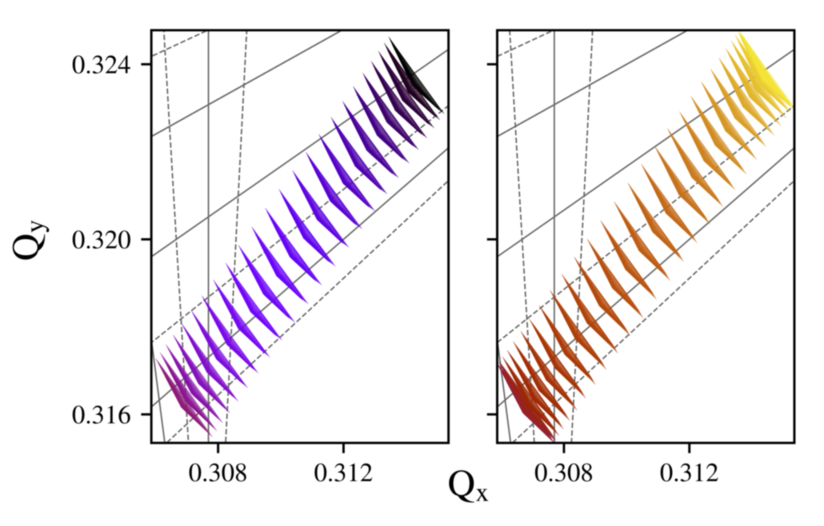}  \\ 
\caption{\label{fig:instanteneous_footprint} The instantaneous footprint in the presence of a tune modulation for the first (left panel) and second (right panel) half of the modulation period. A color code is assigned to the turn number that indicates the first (black) and last (yellow) turns of one modulation period. The resonance diagram (gray lines) is also depicted. }
\end{figure}

The aim of this section is to extend the previous observations to additional degrees of freedom. In 4D and for a distribution of particles, the tune modulation of each particle is translated into a time variation of the tune footprint. To illustrate this effect with tracking simulations, initial conditions forming a polar grid in the configuration space up to 6 $\sigma$ are used, where $\sigma$ stands for one standard deviation assuming a Gaussian distribution and a normalized emittance of 2.5 $\rm \mu m \ rad $. In addition, zero initial transverse momenta are selected. The map consists of a linear rotation, an octupole and a modulated quadrupole. The selected modulation frequency is 20~Hz, which corresponds to a modulation period of \(\rm \approx\)560 turns in the LHC lattice. Such a frequency has been selected to simulate the impact of the synchro-betatron coupling for off-momentum particles in the LHC in the presence of a non-zero chromaticity, as the synchrotron frequency corresponds to approximately 20~Hz. In addition, the linear rotation is divided into eight segments to simulate the eight arcs of the LHC and the transverse coordinates are retrieved after each segment. Combining the data from the eight observation points, the sampling rate increases by a factor of eight compared to the case of a single observation per revolution. An approximation of the instantaneous tune footprint is computed with a window length of 50~turns, which was found to be a good trade-off between time and frequency resolution for the tune determination of each particle. Figure~\ref{fig:instanteneous_footprint} illustrates the instantaneous tune footprint for the first (left panel) and second (right panel) half of the modulation period and a step of 10 turns. The sequential color map represents the time evolution with the start and end of one modulation period shown in black and yellow, respectively. The maximum excursion of the tune footprint from its unperturbed position depends on the selected modulation depth. From the review of the instantaneous footprint, it can be seen that the particles cross several resonances (gray lines) during the modulation period, even if the selected working point is in a resonance-free area of the tune space.

\subsubsection{Frequency Map Analysis with tune modulation}

\begin{figure*}
\subfloat{\subfigimg[width=0.32\textwidth]{\textbf{a)}}{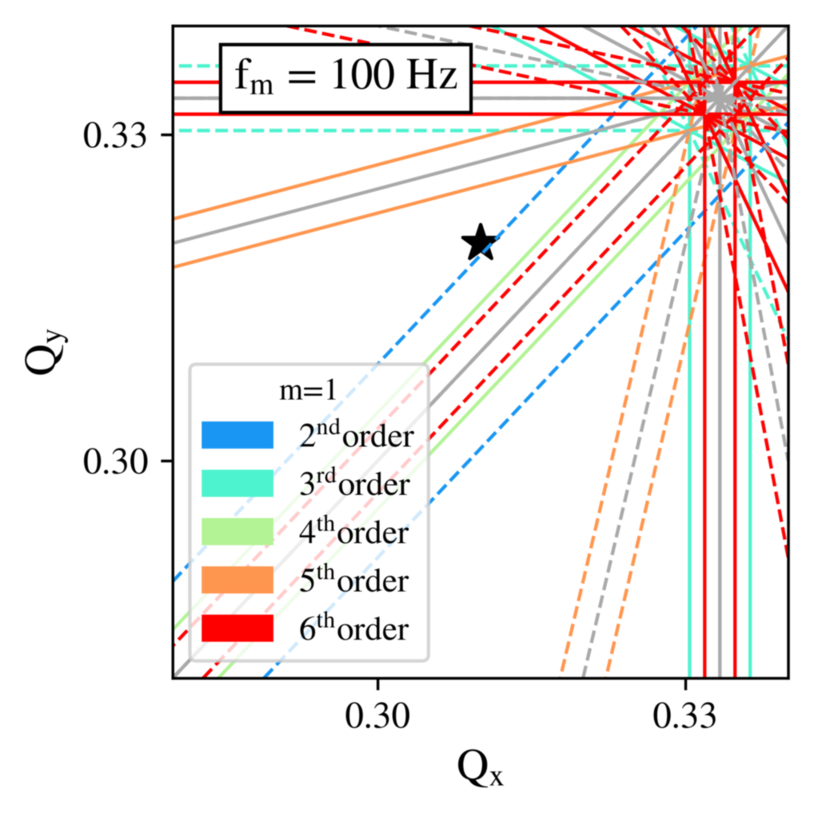} \label{subfig:example_100}}
\subfloat{\subfigimg[width=0.32\textwidth]{\textbf{b)}}{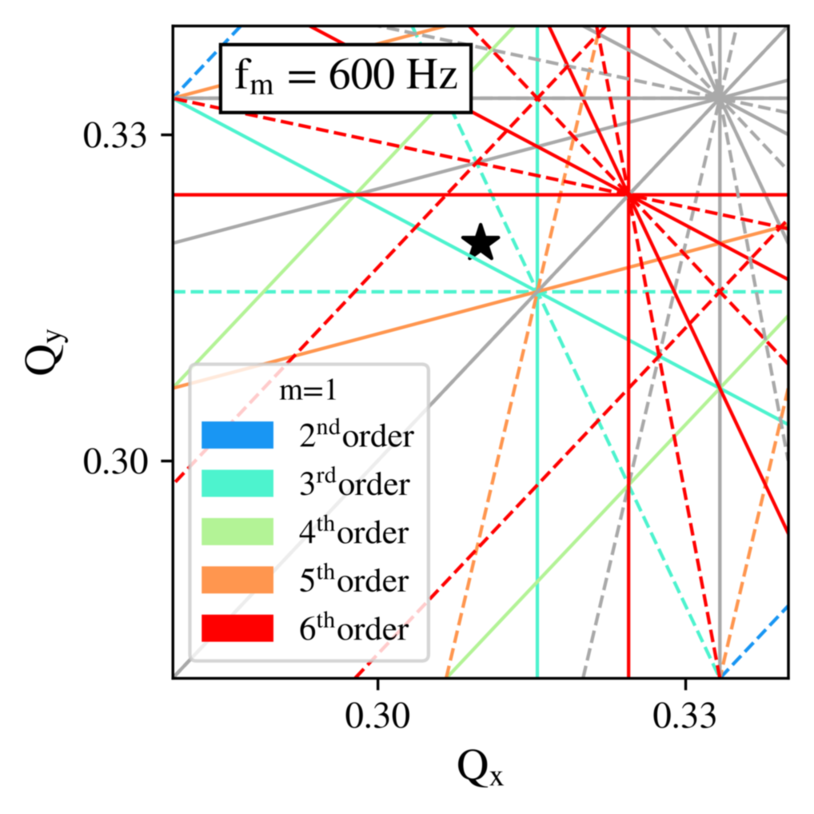} \label{subfig:example_600}}
\subfloat{\subfigimg[width=0.32\textwidth]{\textbf{c)}}{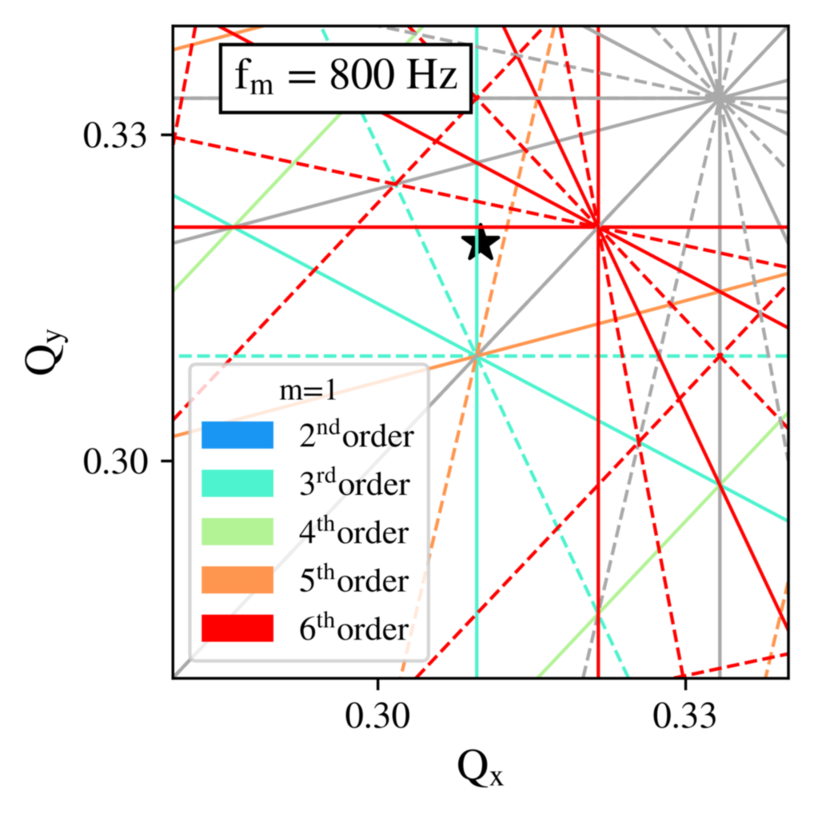} \label{subfig:example_800}}
\caption{\label{example_tune_diagram} The tune diagram with the nominal resonances (gray) and the first-order sideband of the second (blue), third (cyan), fourth (green), fifth (orange) and sixth (red) order resonance for a tune modulation at (a) 100~Hz, (b) 600~Hz and (c) 800~Hz. The working point is also illustrated (black). The solid and dashed lines illustrate the normal and skew resonances, respectively.}
\end{figure*}

In Section~\ref{tune_mod_2D} it was shown that, in the presence of non-linearities, the variation of the tune eventually leads to the excitation of sideband islands. Similarly, averaging the variation of the instantaneous tune footprint over several modulation periods leads to the excitation of sideband resonances in the vicinity of the ones driven by the lattice non-linearities. In particular, in the presence of a tune modulation with a single modulation tune $Q_m=f_{m}/f_{\text{rev}}$, where \(f_{m} \ \text{and} \ f_{\text{rev}}\) are the modulation and revolution frequency, respectively, the resonance diagram is computed as: 

\begin{equation}
k \cdot Q_x + l \cdot Q_y + m \cdot Q_m=n  
\label{eq:resonances}
\end{equation}
where \(k, l, m, n\) are integers and $|k| + |l|$ is the resonance order, while $m$ is the sideband order. Figure~\ref{example_tune_diagram} indicates the working point (black star-shaped marker) and the tune diagram of Eq.~\eqref{eq:resonances} for three different modulation frequencies: a modulation at 100~Hz (Fig.~\ref{subfig:example_100}), 600~Hz (Fig.~\ref{subfig:example_600}) and 800~Hz (Fig.~\ref{subfig:example_800}). The gray lines represent the nominal resonances, while the first-order sideband ($m=1$) of the second (blue), third (cyan), fourth (green), fifth (orange) and sixth (red) order resonance is also depicted ($|k|+|l| \leq 6$). The solid and dashed lines illustrate the normal and skew resonances, respectively. The sidebands are always parallel to the main resonances and they are located in a distance that is proportional to the modulation tune, an effect that is clearly shown in Fig.~\ref{subfig:example_100} with the first-order sideband of the second-order resonance (excited due to linear coupling). As the modulation frequency increases, the sidebands, such as the ones of the second-order resonance (blue) in Fig.~\ref{subfig:example_100}, are driven further away from the working point. At the same time, sidebands of excited resonances that are not in the vicinity of the working point reach the betatron tune spread, such as the sideband of the third-order resonance (cyan) in Fig.~\ref{subfig:example_600} and Fig.~\ref{subfig:example_800}. Subsequently, depending on the selected working point, the lattice non-linearities and the modulation frequency, the sidebands of lower or higher-order resonances might reach the betatron tune spread and increase the particles' tune diffusion.

\begin{figure*}
\subfloat{\subfigimg[width=\columnwidth]{\textbf{a)}}{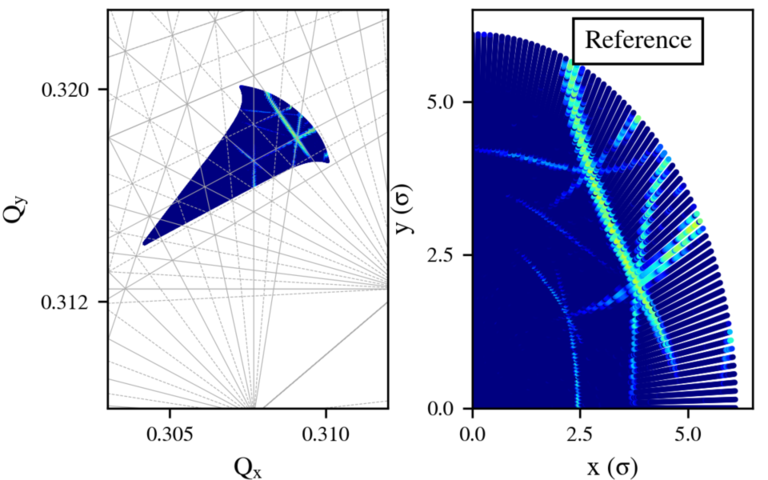} \label{subfig:fma_without}}
\subfloat{\subfigimg[width=\columnwidth]{\textbf{b)}}{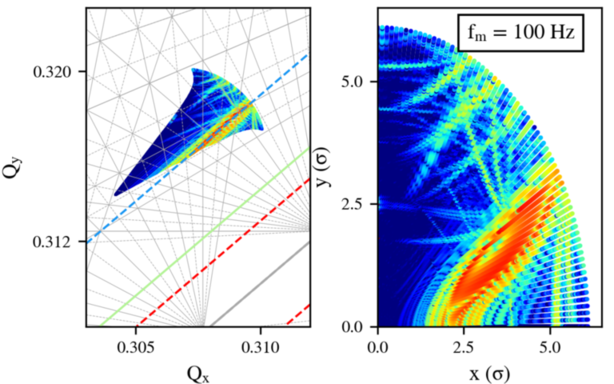} \label{subfig:fma_100}} \\
\subfloat{\subfigimg[width=\columnwidth]{\textbf{c)}}{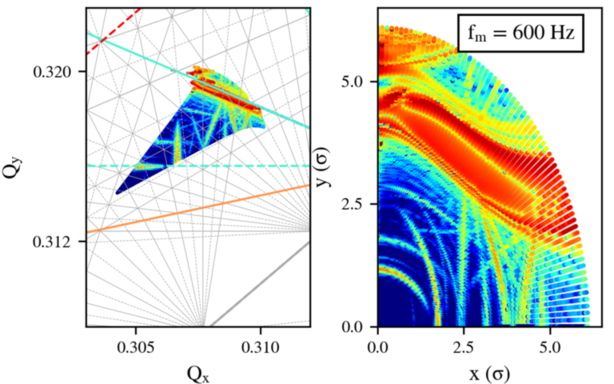} \label{subfig:fma_600}}
\subfloat{\subfigimg[width=\columnwidth]{\textbf{d)}}{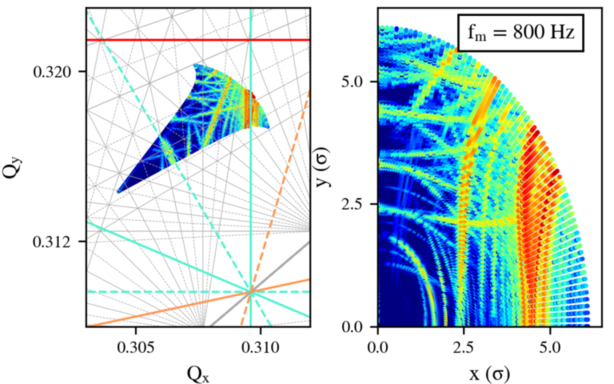} \label{subfig:fma_800}}
\caption{\label{with_without} The FMAs (left panel) of the simplified model (a) in the absence of noise and in the presence of a tune modulation with a frequency at (b) 100~Hz, (c) 600~Hz and (d) 800~Hz, with the initial coordinates in configuration space color-coded with the tune diffusion (right panel). The gray lines depict the nominal resonances, while the colored lines illustrate the first sideband of the resonances up to the sixth order.}
\end{figure*}

The existence and the impact of the sideband resonances described in Eq.~\eqref{eq:resonances} is validated by computing the tune of each particle from the turn-by-turn data that span over several modulation periods. As the aim of the present paper is to study the dynamics of colliding beams that experience strong beam-beam interactions, a 4D beam-beam element is also included in the aforementioned one-turn map. The map of the 4D head-on beam-beam element used in the simulations is described in \cite{herr2016beambeam}, assuming Gaussian and round beams with an intensity of $1.2\cdot 10^{11}$ protons, a normalized emittance of $\rm 2.5 \ \mu m \ rad$ and a $\beta$-function of 100~m. Including such a non-linearity also allows for a more detailed review of the footprint as it leads to the excitation of several resonances and increases the betatron tune spread. The particles are tracked for \(\rm 10^4\) turns, the transverse position and momenta are retrieved turn-by-turn and they are then used to perform the Frequency Map Analysis (FMA) \cite{FMA1, FMA2, FMA3, FMA4}. In particular, the tune of each particle is computed for the first and last 3000 turns, using the NAFF algorithm. The variation of the transverse tunes between the two time intervals reveals information concerning its tune diffusion. Figure~\ref{with_without} shows the tune determination in the second time span (left panel) and the initial configuration space (right panel), color-coded with the logarithm of the tune diffusion for four studies: in the absence of a tune modulation, which is used as a reference (Fig.~\ref{subfig:fma_without}), and in the presence of a tune modulation at 100~Hz (Fig.~\ref{subfig:fma_100}), 600~Hz (Fig.~\ref{subfig:fma_600}) and 800~Hz (Fig.~\ref{subfig:fma_800}). As a reference, the typical values of the tune diffusion extend from $10^{-7}$ (blue) to $10^{-3}$ (red). The review of the frequency maps shows that for a tune modulation at 100~Hz (Fig.~\ref{subfig:fma_100}) the first-order sideband of the second-order resonance (blue) has a clear impact on the tune diffusion of the particles. This observation underlines that there is a good agreement between Eq.~\eqref{eq:resonances} and the tracking results. The increase of the modulation frequency at 600~Hz results in an important increase of the tune diffusion due to the first sideband of the third-order resonance (cyan). Although lower-order resonances are more critical, the configuration space (right panel) shows that this resonance affects a large portion of the phase space, an effect that explains the more critical impact compared to the modulation at 100~Hz. Finally, the same sideband affects the footprint during a modulation at 800~Hz, but due to the position of the sidebands, a lower impact is observed.

Overall, in the presence of a tune modulation, the impact on the particles' tune diffusion depends on the lattice non-linearities, the position of the sidebands, the sideband and resonance order (lower orders have a more critical impact), the working point and the actions of the affected particles. If the sidebands reach high amplitude particles rapid losses are observed in a limited amount of turns. Therefore, tune modulation effects introduce a frequency-dependent diffusion mechanism, the impact of which significantly depends on the selected working point and the non-linearities of the lattice. This fact underlines that some modulation frequencies are more critical for the beam performance than others for the selected working point. Contrary to modulated dipolar perturbations, where a significant impact on the particle's motion is observed, to first order, when the frequencies of the spectral components are in the proximity of the betatron tune, power supply ripples in the quadrupoles can affect the distribution even if the modulation frequency is not in the vicinity of the working point.

\section{Tune modulation in the HL-LHC}
\label{sec:HL-LHC}

In the HL-LHC, luminosity levelling techniques are required to achieve a constant luminosity of \(\rm 5\times 10^{34} \ cm^{-2}s^{-1}\), as envisaged in the nominal scenario \cite{HL_LHC_scenarios, Apollinari:2284929}. The luminosity degradation, resulting from the intensity reduction from \(2.2 \cdot 10^{11}\) to \(1.2 \cdot 10^{11}\) protons per bunch due to the proton-proton collisions, will be compensated by reducing the beam size at the two high luminosity experiments, CMS and ATLAS, from \(\beta^*\)=64~cm to 15~cm with the Achromatic Telescopic Squeezing (ATS) scheme \cite{ATS}. In the context of these studies, the machine configuration at the end of the levelling is the most critical as the increase of the maximum \(\beta\)-functions in the inner triplet will act as an amplification factor for the noise in the quadrupoles. Therefore, simulations with power supply ripples in the inner triplet left and right of the two IPs are conducted for the simulated parameters shown in Table~\ref{tab:simulation_parameters}. As a first step, arbitrary modulation amplitudes and frequencies are employed to illustrate the impact of such a mechanism on the particles' motion either in 5D, i.e., without considering the synchrotron oscillations but with off-momentum particles or in 6D to depict the combined effect of the gradient and the synchrotron modulation. From this parametric investigation, the dependence on the modulation frequency is presented in terms of tune diffusion, intensity evolution and DA. The larger sensitivity to specific modulation frequencies is explained with frequency maps and a modulation depth threshold is defined as a function of the frequency.

\begin{table}
\caption{The HL-LHC simulated parameters at the end of the luminosity levelling at top energy.}
\label{tab:simulation_parameters}
\begin{ruledtabular}
\begin{tabular}{lc}
\textrm{Parameters (unit)}& \textrm{HL-LHC v1.3 (values)}\\
\colrule
Beam energy (TeV) &  7\\
Bunch spacing (ns) &  25\\
Bunch length (m) & 0.075 \\
Betatron tunes (\(Q _x, Q_y\))  &  (62.31/62.315\footnote{Nominal/optimized working point.}, 60.32) \\
Normalised emittance (\(\rm \mu m \ rad\)) & 2.5\\
Chromaticity  & 15\\
Octupole current (A)  & -300\\
Bunch population (protons)  & 1.2e11\\
IP1/5 Half crossing angle (\(\rm \mu rad\))  &  250\\
IP1/5 \(\beta^*\) (cm) & 15 \\
Crabbing angle (\(\rm \mu rad\))& 190 \\
\end{tabular}
\end{ruledtabular}
\end{table} 

Of particular importance is the fact that the quadrupoles in the inner triplet are powered by switch-mode power supplies \cite{SM_PC}. In contrast to thyristor commutated technology, the voltage tones that are anticipated in the noise spectrum, in this case, are the switching frequency of the power supply and its harmonics, as well as a few low order 50~Hz harmonics, namely 50, 150, 300 and 600~Hz, without excluding the existence of others \cite{Gamba_2017}. It must be noted that the envisaged switching frequencies of the main circuits in the triplet lie in the regime of 50-200~kHz \cite{Gamba_2019}. Such high frequencies are not expected to perturb the beam motion as they will be strongly attenuated by the shielding effect of the beam screen \cite{Martino}. Therefore, the following studies mainly focus on the potential impact of the low order harmonics in the spectrum or low switching frequencies in the trim circuits. As a realistic voltage spectrum is not presently available, the results of the parametric simulations are compared to the power supply specifications \cite{Gamba_2017}. The specifications provide the maximum tolerated output voltage for an extended frequency bandwidth, without defining the actual amplitude and the voltage tones in the noise spectrum of the triplet. These values, in combination with the inductance and the current of the circuits, are used to compute the expected tune modulation from the four quadrupoles and the three trims left and right of the two IPs. Furthermore, the transfer function that converts the voltage ripples into variations of the magnetic field is approximated by a simple RL circuit and the beam screen attenuation \cite{Martino} is not included, an approach which leads to the overestimation of the modulation depths in these studies.

\begin{figure}
\includegraphics[width =\columnwidth]{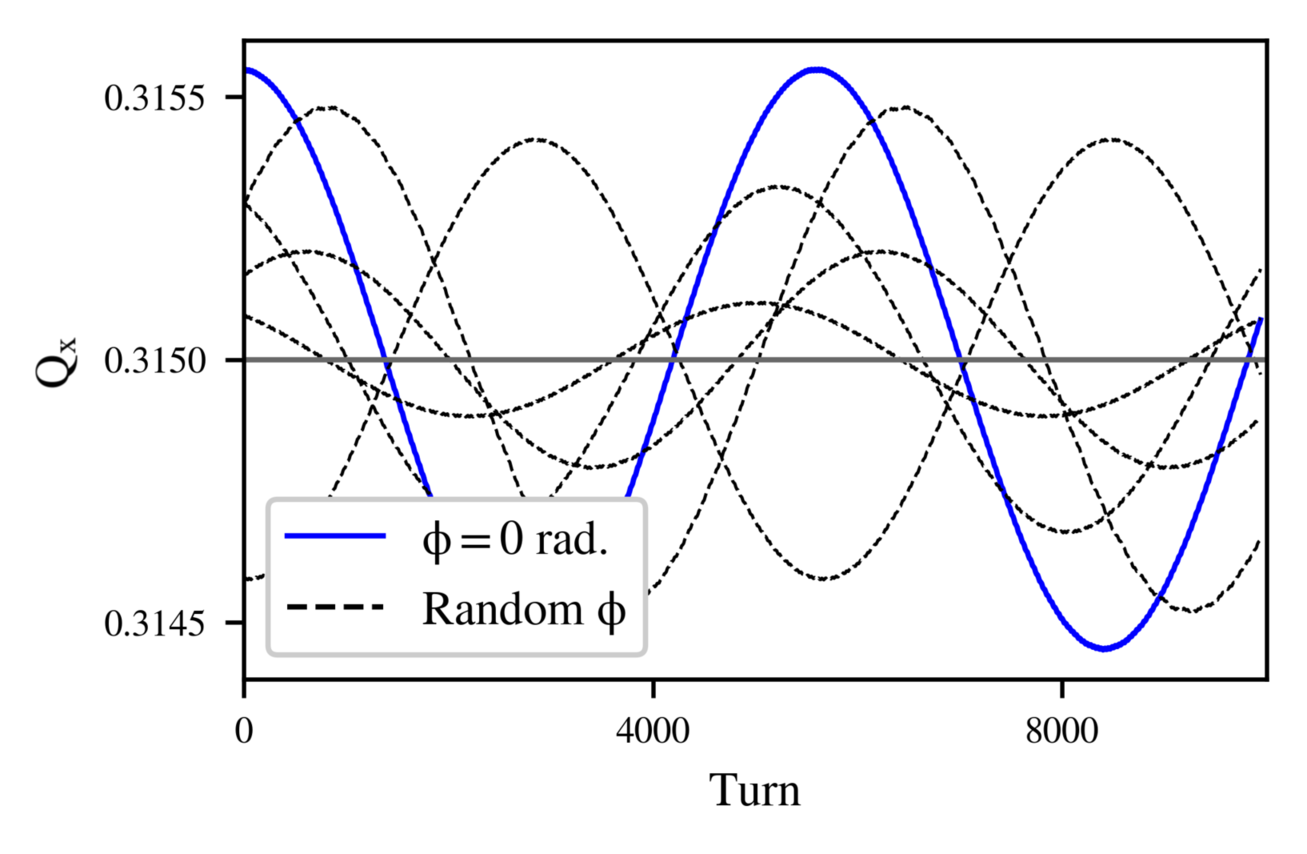} 
\caption{\label{fig:hl_phase} The horizontal instantaneous tune in the presence of a low-frequency modulation with a zero (blue) and a random (black) phase of the noise in the inner triplet quadrupoles left and right of IP1 and 5. The gray line represents the horizontal tune in the absence of a modulation.}
\end{figure}

The simulations are performed using the single-particle symplectic tracking code, SixTrack \cite{sixtrack, sixtrack2}. Only one beam is tracked around the lattice, that corresponds to Beam 1, and the weak-strong approximation is used for the beam-beam effects, as the charge distribution of the strong beam is not varied. The normalized strengths of Q1, Q2a, Q2b and Q3 in the IRs left and right of IP1 and 5 are modulated with a sinusoidal function. The amplitude of the function defines the maximum variation of the strength while the polarity of each magnet is preserved. The absolute value of the amplitude is assumed to be equal for all the quadrupoles, without however leading to the same modulation depth due to variations of the \(\beta-\)functions. The phase of the injected noise is considered equal to zero for all the circuits. This approximation does not correspond to a realistic scenario, as the switching clocks of the power supplies are not synchronized. Nevertheless, such a configuration is selected as it maximizes the overall impact of the noise. In fact, due to the anti-symmetric powering of the triplets left and right of the two IPs, the modulation of the circuits accumulates when equal phases are considered. To illustrate this effect, a single particle is tracked in the HL-LHC lattice for \(\rm 10^4\) turns. The simulations are performed in 5D to disentangle the contribution of the synchrotron motion from the power supply ripples. A low-frequency modulation is selected so that an approximation of the instantaneous tune can be derived from the turn-by-turn position information. The horizontal tune is computed with a sliding window of 50~turns with a step of 10 turns between intervals. Figure~\ref{fig:hl_phase} presents the tune evolution for equal (blue) and random (black) phases in the circuits. In the latter case, the randomly selected phases act as a compensation scheme for the overall modulation compared to the former. The unperturbed horizontal tune is also depicted (gray). Several studies are performed with random phases to validate this observation. Specifically, due to the betatron phase advance between the right and left circuits around the IPs, the modulation is minimized when their phase difference approaches \(\Delta \phi = \rm \pi\). Thereby, simulations show that there is a potential benefit of phasing conveniently the clocks of the power supplies. The results of the above simulations are presented and discussed in the next sections.

\subsection{Noise spectrum with a single tone}

\subsubsection{Frequency Map Analysis in 5D with noise}

\begin{figure*}
\subfloat{\subfigimg[width=0.49\linewidth]{\textbf{a)}}{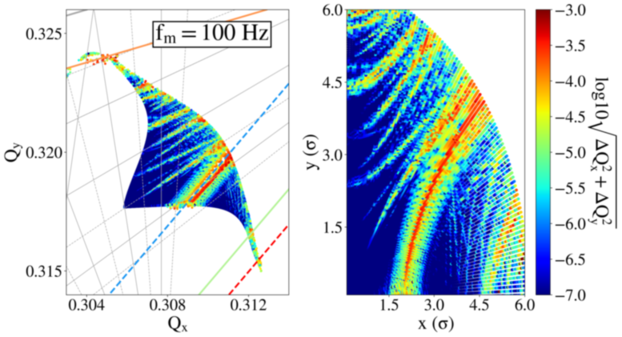} \label{subfig:hl_fma_100}}
\subfloat{\subfigimg[width=0.49\linewidth]{\textbf{b)}}{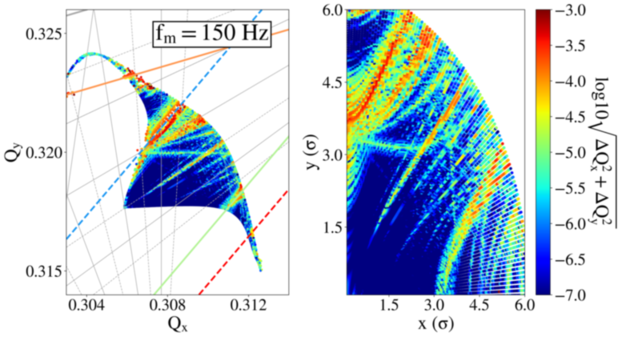} \label{subfig:hl_fma_150}} \\
\subfloat{\subfigimg[width=0.49\linewidth]{\textbf{c)}}{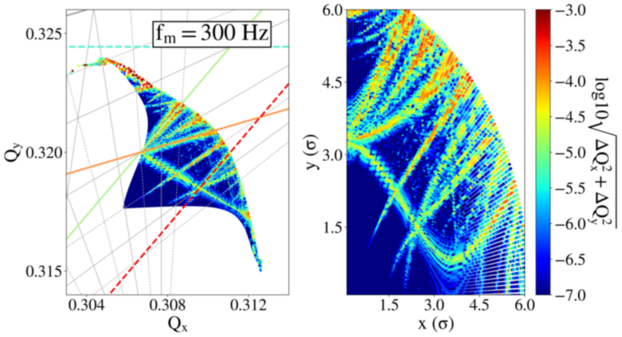} \label{subfig:hl_fma_300}}
\subfloat{\subfigimg[width=0.49\linewidth]{\textbf{d)}}{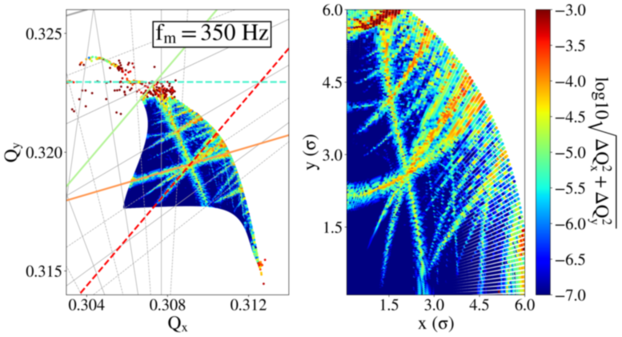} \label{subfig:hl_fma_350}} \\
\subfloat{\subfigimg[width=0.49\linewidth]{\textbf{e)}}{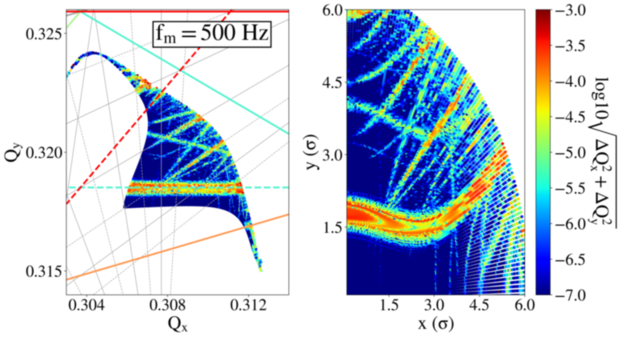} \label{subfig:hl_fma_500}}
\subfloat{\subfigimg[width=0.49\linewidth]{\textbf{f)}}{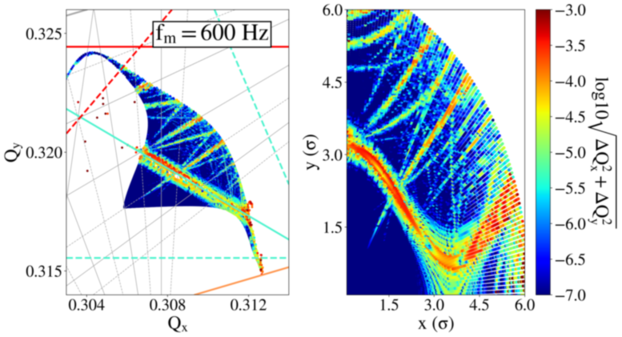} \label{subfig:hl_fma_600}}
\caption{\label{fig:hl_fmas} The 5D frequency map (left panel) and the initial configuration space (right panel) in the presence of a tune modulation in the inner triplet left and right of IP1 and 5 in the HL-LHC lattice for the nominal working point. The modulation depth is \(\Delta Q = \rm 5.5 \cdot 10^{-5}\) and the modulation frequency is (a) 100~Hz, (b) 150~Hz, (c) 300~Hz, (d) 350~Hz, (e) 500~Hz and (f) 600~Hz. A color code is assigned to the logarithm of the tune diffusion. The nominal (gray) lines and the first-order sideband of the second (blue), third (cyan), fourth (green), fifth (orange) and sixth (red) order resonance are indicated.}
\end{figure*}

Similarly to the simplified model of Section~\ref{henon}, the FMAs are computed to depict the excitation of sideband resonances in the HL-LHC case due to noise effects. A distribution of particles is tracked in 5D for \(\rm 10^4\) turns in the lattice at the nominal working point. The initial conditions form a polar grid in the configuration space and zero initial transverse momenta are selected. The grid consists of 99 angles and a radius extending from 0.1 to 6.1~\(\rm \sigma\) and a step of 1~\(\rm \sigma\). Longitudinally, the particles are located at 3/4 of the bucket height. The turn-by-turn data are divided into two groups, consisting of the first and last 3000 turns, respectively. The tune diffusion of the particles is computed by comparing the variation of their tune between the two time intervals. Figure~\ref{fig:hl_fmas} depicts the FMAs (left panel) and the initial configuration space (right panel), color-coded with the logarithm of the tune diffusion, for several modulation frequencies and equal depths. The gray lines represent the nominal resonances, while the colored lines illustrate the first sideband ($m=1$) of the second (blue), third (cyan), fourth (green), fifth (orange) and sixth (red) order resonance ($|k| + |l| \leq 6$).

For a constant modulation depth, the impact of the tune modulation depends on the working point and the modulation frequency. For a modulation at 100~Hz (Fig.~\ref{subfig:hl_fma_100}) and 150~Hz (Fig.~\ref{subfig:hl_fma_150}) the first sideband of the second-order resonance (blue) leads to an increase of the tune diffusion. For a modulation at 300~Hz (Fig.~\ref{subfig:hl_fma_300}), sidebands of higher-order resonances (fourth to sixth) are in the vicinity of the betatron tune. A modulation at 350~Hz (Fig.~\ref{subfig:hl_fma_350}) leads to rapid losses due to the fact that the first sideband of the third-order resonance (cyan) reaches high-amplitude particles and overlaps with the nominal resonances (gray lines), as well as with the first sideband of the fourth-order resonance (green). Finally, for a modulation at 500~Hz (Fig.~\ref{subfig:hl_fma_500}) and 600~Hz (Fig.~\ref{subfig:hl_fma_600}) the excitation of the first sideband of the third-order resonance (cyan) is visible and the review of the configuration space shows that the core of the distribution is mainly affected. From the frequency maps, it is observed that there is a sensitivity to some modulation frequencies as the same modulation depth leads to a higher diffusion increase. As previous studies have reported \cite{HL_LHC_rip}, for the working point and the non-linearities of the HL-LHC lattice, amongst the low-order harmonics that are anticipated in the spectrum (50,150,300 and 600~Hz), a higher sensitivity to 300 and 600~Hz is present. This is due to the fact that multiple sidebands are reaching the footprint. This effect is not observed with a modulation at 50 and 150~Hz as only the first sideband of the second-order resonance is in the vicinity of the betatron tune and it affects a small portion of the phase space.

\subsubsection{Frequency Map Analysis in 6D}

\begin{figure}
\includegraphics[width =\columnwidth]{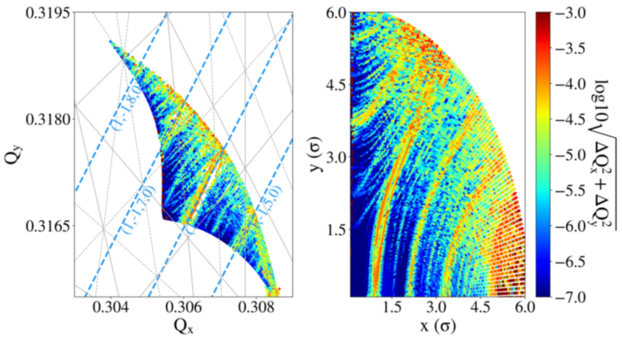} 
\caption{\label{fig:hl_fma_6D} The 6D FMA with a chromaticity equal to 5 and for a momentum deviation of all particles at 3/4 of the bucket height. The blue lines depict the sidebands ($m \leq 8$) of the second-order resonance.}
\end{figure}

In a similar way, the coupling of the synchrotron and the betatron motion in the presence of a non-vanishing chromaticity leads to the variation of the instantaneous footprint with a modulation frequency equal to the synchrotron frequency ($\approx$20~Hz). As the synchrotron frequency in the LHC is much lower than the ripple frequencies under consideration, the tracking is extended to \(\rm 2 \cdot 10^4 \) turns to average over several modulation periods. The same distribution of particles is tracked in 6D in the nominal HL-LHC lattice, i.e., without injecting noise in the triplet. The momentum deviation of all the particles is equal to 3/4 of the bucket height. It is important to note that, as shown in Section~\ref{henon}, the modulation index of each particle with a chromaticity of 15, a relative momentum deviation of $\rm 27\cdot 10^{-5}$ and a synchrotron frequency at \(\rm 20~Hz\) exceeds the critical value of 1.5. In this case, as presented in the previous section, due to the appearance of strong sidebands, the NAFF algorithm returns the frequency determination of the sideband and not the one of the betatron tune. Therefore, to illustrate the frequency map in 6D the chromaticity is reduced to 5. The turn-by-turn data obtained by tracking are divided into two consecutive intervals of \(\rm 10^4 \) turns and the tune diffusion of each particle is computed. Figure~\ref{fig:hl_fma_6D} demonstrates the 6D FMA (left panel) and the initial configuration space (right panel). The frequency map shows that high order synchrotron sidebands ($m\leq 8$) of the second-order resonance (blue) are excited. Although considering constant momentum deviations for all the particles in the distribution is essential to compute frequency maps, experimentally the initial longitudinal distribution extends over the whole bucket area. As the momentum deviation is not constant across all particles, the variation of the modulation index leads to the reduction of the average modulation. To this end, the next section focuses on the investigation of the intensity evolution in a more realistic configuration.

\subsubsection{Simulations of beam losses}

Studying the intensity evolution requires the tracking of a 6D matched distribution with momentum deviations that span over the whole bucket height. Due to the excitation of additional resonances from the slow (synchrotron motion) and fast (power supply ripple) modulation, particles at the tails of the distribution diffuse and will eventually be lost. A detailed representation of the high amplitude particles is achieved by first, overpopulating the tails and then, assigning weights to the particles according to their initial position. Specifically, the initial conditions form a uniform 4D round distribution in the transverse plane that extends up to 6 \(\rm \sigma \) both in the configuration and the trace space. Longitudinally, the particles are uniformly placed in the bucket height extending up to its limit. Overall, \(\rm 9\cdot 10^4 \) particles are tracked for \(\rm 10^6 \) turns and the turn-by-turn position measurements are retrieved every \(\rm 10^3 \) turns. Depending on the initial position of the particles in the transverse and longitudinal plane, a weight is assigned to each particle in the post-processing analysis. The weight defines the importance of each particle in the computations of the intensity. It is determined from the Probability Density Function (PDF) of the simulated distribution, which in this case is Gaussian in all planes. Instead of directly tracking a Gaussian distribution, an important number of particles are located at the tails of the distribution, to which a lower weight is assigned compared to the ones at the core. In this way, their contribution to the determination of the losses is less significant. Furthermore, a mechanical aperture is defined in the post-processing at 5 \(\rm \sigma\). Particles reaching this limit are considered lost and the corresponding weights are set to zero.

\begin{figure*}
\subfloat{\subfigimg[width=\columnwidth]{\textbf{a)}}{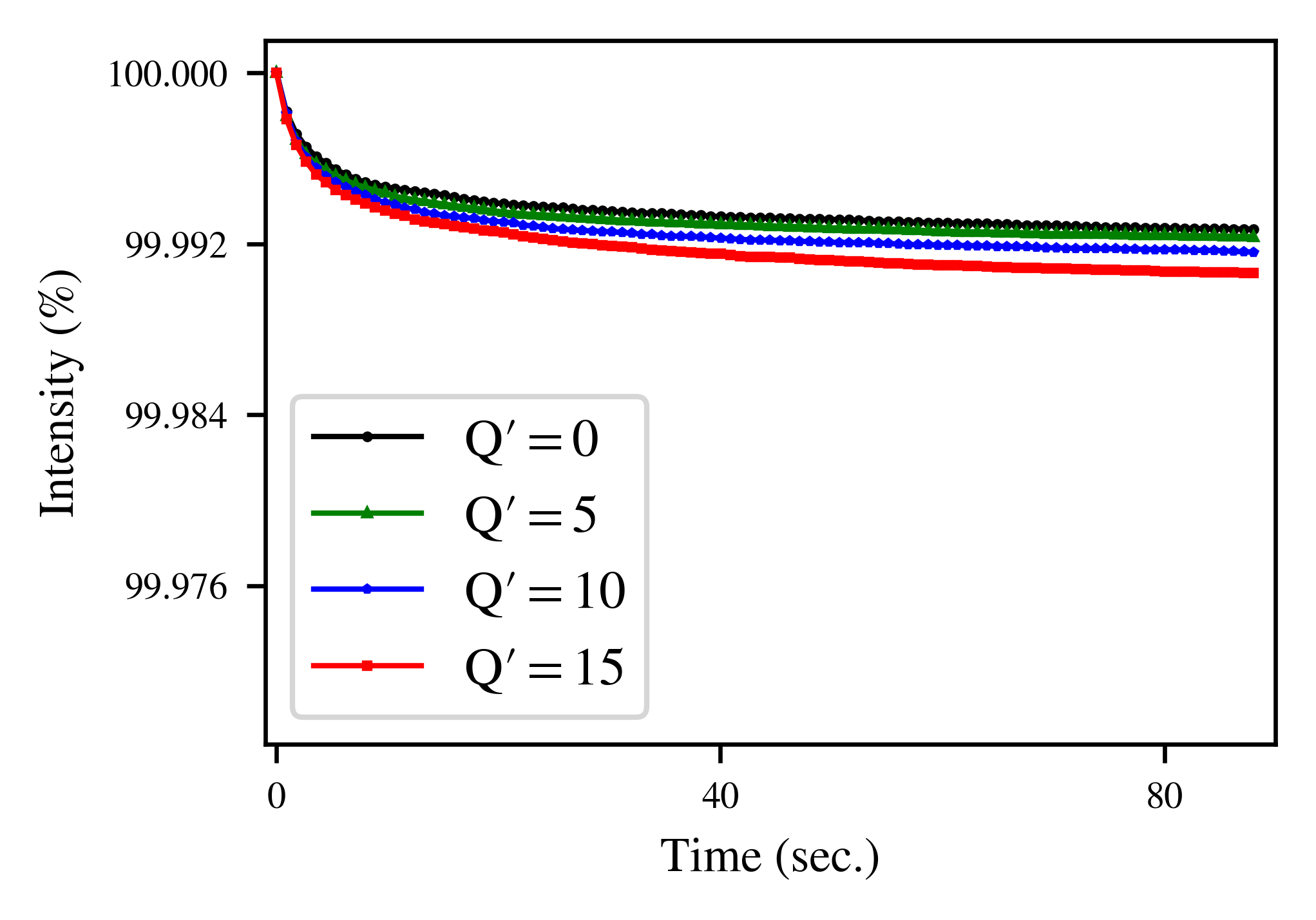} \label{subfig:losses_chroma}} 
\subfloat{\subfigimg[width=\columnwidth]{\textbf{b)}}{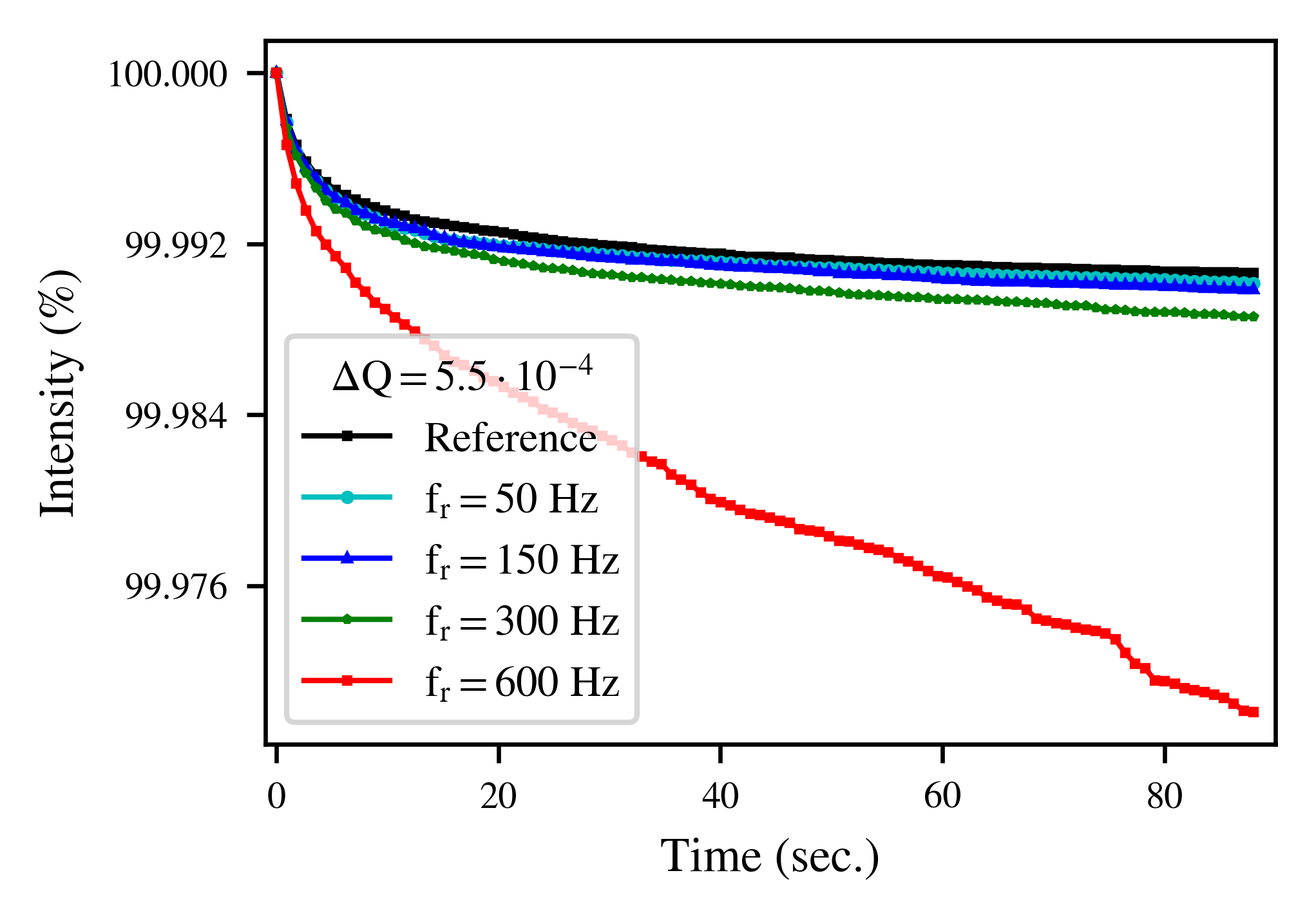} \label{subfig:losses_ripples}} 
\caption{\label{fig:hl_lifetime_noise_chroma} (a) Intensity evolution in the HL-LHC lattice for \( Q'=\rm 0 \) (black), \(Q'=\rm 5 \) (green), \(Q'=\rm 10\) (blue) and \(Q'=\rm 15\) (red). (b) Intensity evolution for \(\rm Q'=15\) without noise (black) and in the presence of a tune modulation with \(\Delta Q = \rm  5.5 \cdot 10^{-5}\) at 50~Hz (cyan), 150~Hz (blue), 300~Hz (green) and 600~Hz (red).}
\end{figure*}

Figure~\ref{fig:hl_lifetime_noise_chroma} shows the intensity evolution as computed from the weighted distributions. First, the chromaticity and thereby, the modulation depth is increased in steps (Fig.~\ref{subfig:losses_chroma}) from 0 to 15, which leads to a lifetime reduction. The intensity evolution shows that the intensity degradation scales roughly linearly with the chromaticity increase. For a constant chromaticity equal to 15 in both planes, the frequency of the tune modulation induced by power supply ripples is varied for a constant modulation depth (Fig.~\ref{subfig:losses_ripples}). The study focuses on the low-frequency tones that are expected to be present in the power supply (50, 150, 300 and 600~Hz) and that may perturb the beam motion. Although, for a constant modulation depth, the modulation index decreases with increasing frequency, the impact on the intensity is much more severe for 600~Hz compared to 50, 150 and 300~Hz. The beam lifetime is computed in each case with an exponential fit which yields a reduction from $\approx$ 1100~h for a modulation frequency at 50~Hz to $\approx$750~h for 300~Hz and $\approx$130~h when the frequency is increased to 600~Hz. As shown from the FMAs, the tune modulation introduces a frequency-dependent mechanism of tune diffusion, which, for the selected working point, renders these frequencies the most critical in terms of losses as with the same modulation depth a more significant impact on the intensity is observed.

\subsubsection{Noise thresholds with parametric Dynamic Aperture scans}

\begin{figure*}
\includegraphics[width = \textwidth]{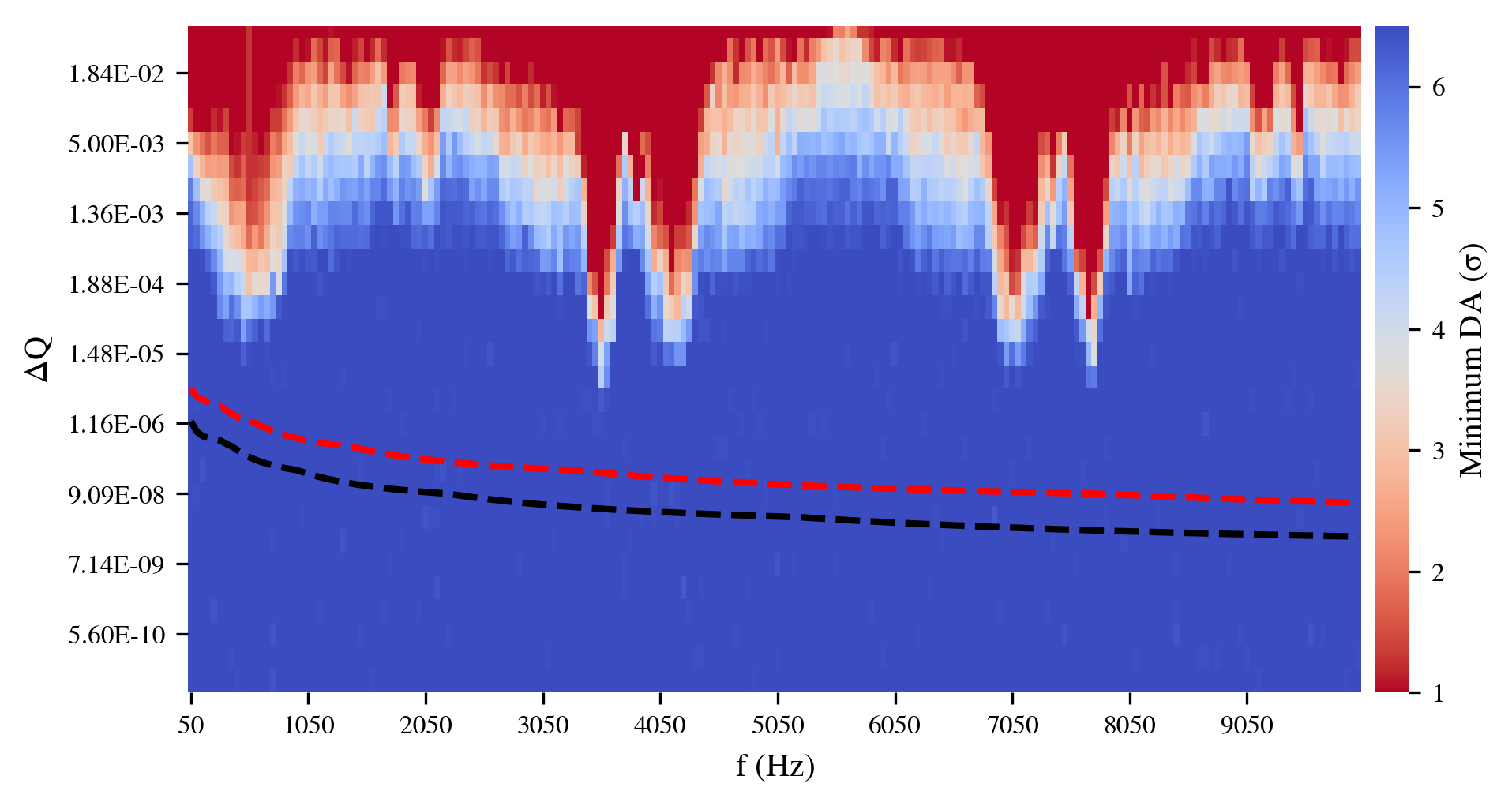} 
\caption{\label{fig:hl_heat} Modulation frequency as a function of modulation depth in the presence of noise in the inner triplet quadrupoles of the HL-LHC. The color code indicates the minimum DA. The red and black lines represent the sum and the root mean square of the contribution from all the circuits, respectively, computed from the maximum output voltage as defined in the power supply specifications.}
\end{figure*}

As the presence of voltage tones with frequencies beyond the ones already considered cannot be excluded, the DA dependency on the modulation frequency is explored. To this end, parametric 6D simulations are performed to estimate the impact of various modulation frequencies and depths. In the transverse configuration space, the initial distribution consists of five angles and a radius that extends from two to ten \(\rm \sigma\) with a step of two \(\rm \sigma\). In the longitudinal plane, all particles are placed at 3/4 of the bucket height. The duration of the tracking is \(\rm 10^6\) turns, which corresponds to \(\rm \approx\)90 seconds in operation. The working point is set to its nominal \((Q_x, Q_y) = \rm  (62.31, 60.32)\) and not the DA optimized \((Q_x, Q_y) = \rm  (62.315, 60.32)\) value \cite{DA_optimize}. For each study, a different combination of the excitation frequency and amplitude is selected in order to conduct a scan in the tune modulation parameter space. The frequency range spans over all 50~Hz harmonics up to 10~kHz. For each frequency, the amplitude of the excitation is increased and the total modulation depth due to the contribution of all the circuits is computed from the optics using MAD-X \cite{madx}. The minimum DA across all the angles in the configuration space is computed. Figure~\ref{fig:hl_heat} presents the modulation frequency as a function of the modulation depth. Each point on the plot corresponds to an individual study and a color code is assigned to the minimum DA. The parametric scan defines the modulation depth threshold for each modulation frequency beyond which a DA reduction is foreseen (white to red). As an average, a maximum limit in the order of \(\rm 10^{-4}\) in the modulation depths can be defined; however the strong dependency on the modulation frequency is evident and, as expected from the frequency maps, there are regimes with increased sensitivity to power supply ripples, which is further investigated in the following section. Furthermore, these simulations include the combined effect of ripples in the triplet and high chromaticity for large momentum deviations. 

Next, the modulation depth is computed from the maximum output voltage as defined in the power supply specifications. For each triplet, the contribution of the four main power supplies and the three trim converters are considered. The sum (red dashed) and the root mean square (black dashed) modulation depth from the contribution of all the circuits are computed. It must be highlighted that these values correspond to the maximum acceptable and not the realistic voltage. The attenuation from the beam screen is not considered from which a further reduction of the modulation depth is expected for frequencies above \(\rm \approx\) 100~Hz \cite{Martino}. In addition, summing the contribution of all the triplets corresponds to a scenario where the phase of the power supply noise is synchronous and thus, overestimates the modulation depth even further. A comparison between the modulation depth from the specifications and the threshold defined by simulations through the scan of individual frequency indicates that there is a difference of several orders of magnitude between the two. Therefore, the much larger tolerances defined by the DA simulations suggest that, by considering individual tones, the ripples in the inner triplet, combined with large chromaticity values, will not pose a limitation for the beam performance.

\subsubsection{A review of the high-sensitivity power supply noise regimes}

\begin{figure*}
\subfloat{\subfigimg[width=\columnwidth]{\textbf{a)}}{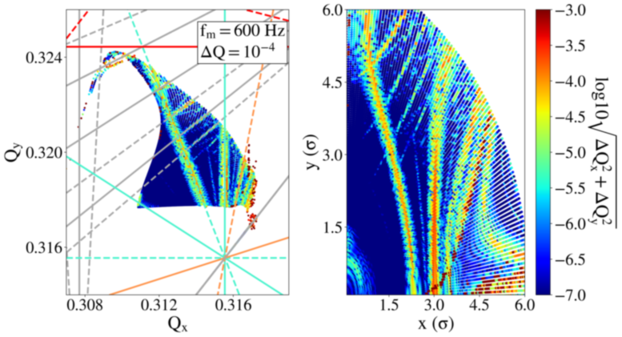} \label{subfig:fma_600_1}} 
\subfloat{\subfigimg[width=\columnwidth]{\textbf{b)}}{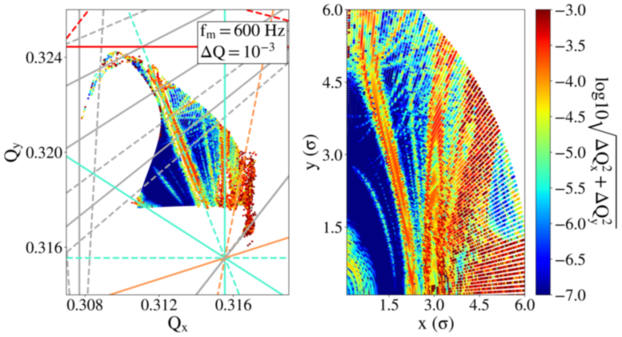} \label{subfig:fma_600_2}} 
\caption{\label{fig:hl_fma_600Hz_vs_depth} The 5D frequency maps for a tune modulation at 600~Hz with (a) \(\Delta Q =\rm 10^{-4}\) and (b) \(\Delta Q = \rm 10^{-3}\). The gray lines represent the nominal resonances, while the cyan lines the first sideband of the third-order resonance.}
\end{figure*}

Based on the heat-map of Fig.~\ref{fig:hl_heat}, a more detailed review is performed around the frequency regimes with a higher sensitivity to power supply ripples. First, to illustrate the repercussions of the modulation depth increase, two sets of noise parameters are selected, corresponding to a slight and a significant DA reduction, respectively. For both cases the modulation frequency is 600~Hz and the optimized working point is selected. Figure~\ref{fig:hl_fma_600Hz_vs_depth} depicts the 5D FMA for \(\Delta Q = \rm 10^{-4}\) (Fig.~\ref{subfig:fma_600_1}) and \(\Delta Q =\rm 10^{-3}\) (Fig.~\ref{subfig:fma_600_2}). The first-order sideband of the third (cyan), fifth (orange) and sixth (red) order resonance is depicted. The review of the frequency maps demonstrates that the critical impact on the DA when increasing the modulation depth is observed due to, first, the increase of the sideband resonance strength and, second, the appearance of higher-order sidebands.

\begin{figure}
\subfloat{\subfigimg[width=0.98\columnwidth]{\textbf{a)}}{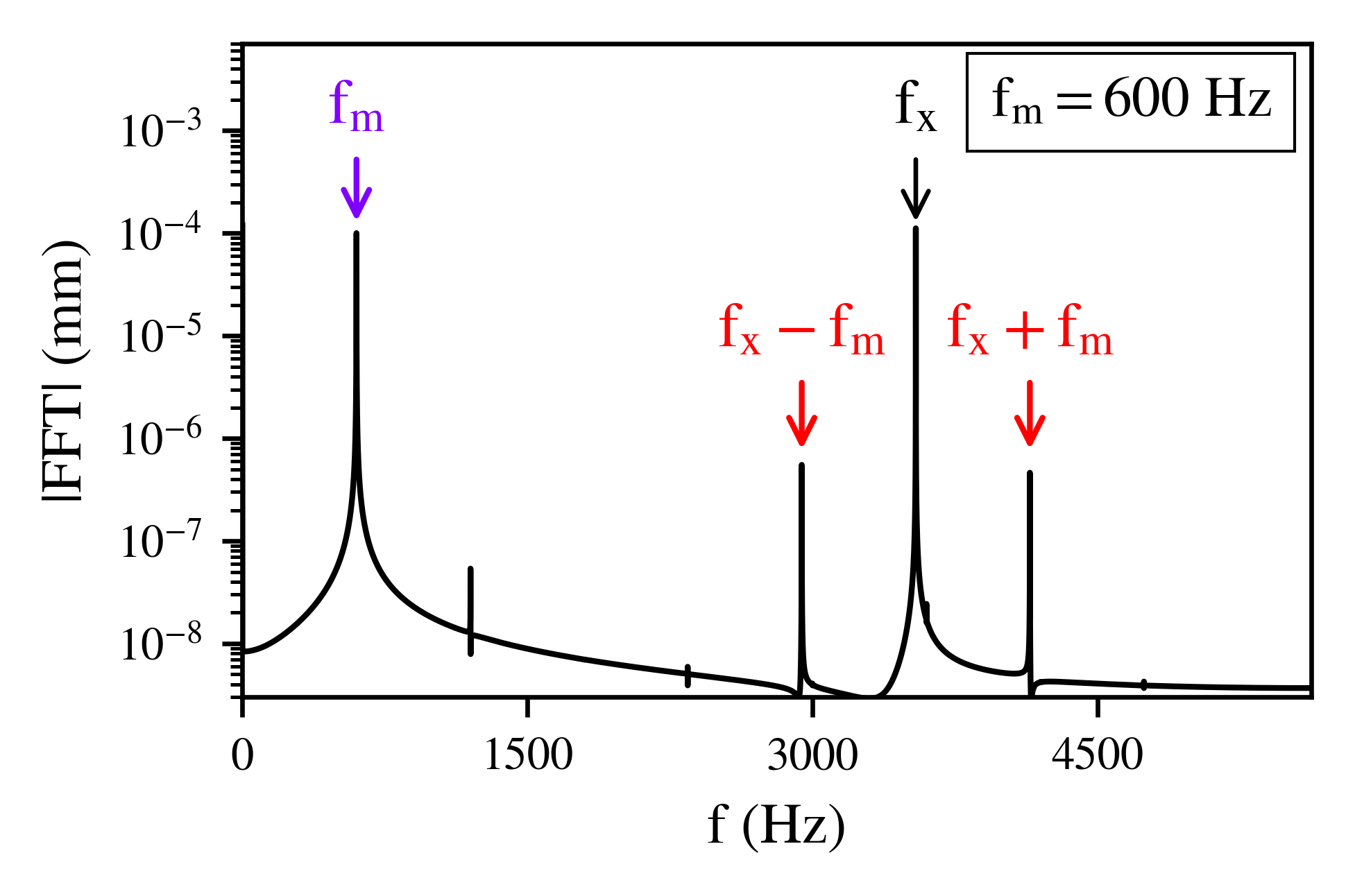} \label{subfig:fft_600}} \\
\subfloat{\subfigimg[width=0.98\columnwidth]{\textbf{b)}}{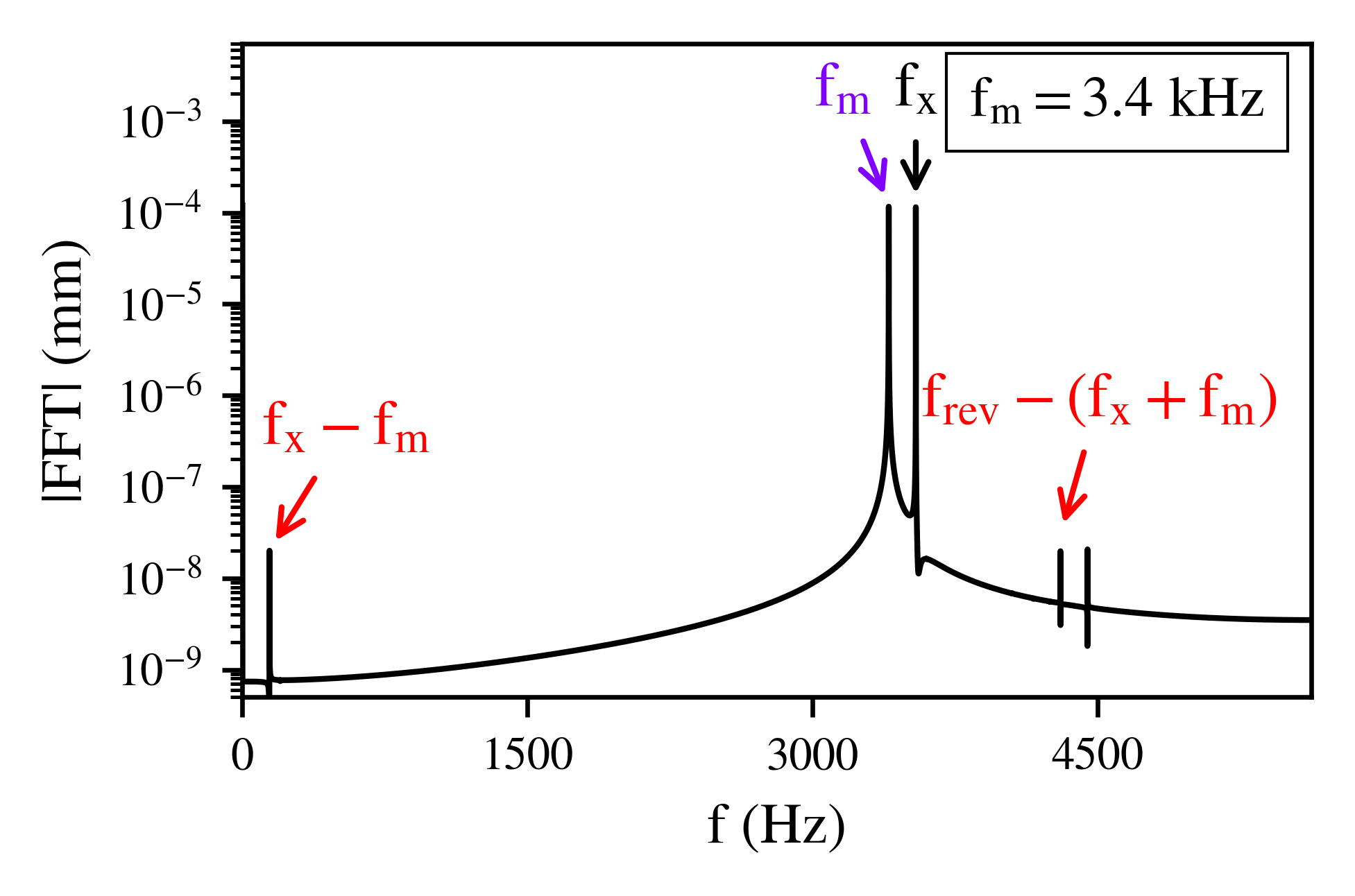} \label{subfig:fft_3400}} 
\caption{\label{fig:hl_FFT} FFT for a tune modulation at (a) 600~Hz and (b) 3.4~kHz that depicts the sidebands (red) around the betatron tune (black) and the dipolar excitation (purple) due to feed-down.}
\end{figure}

Secondly, a higher sensitivity is observed around the regime of the betatron frequency (\(f_x\approx\)3.48~kHz) and its alias, i.e., the folding of the betatron frequency around the revolution frequency (\(f_{\text{rev}}-f_x\approx\)7.76~kHz). This observation is attributed to the dipolar effect of the triplets through feed-down. In particular, as the particle trajectories are not aligned to the magnetic center of the quadrupoles, an orbit modulation is also observed. This is demonstrated by tracking a single particle in the HL-LHC lattice under the influence of a tune modulation at 600~Hz and 3.4~kHz. From the turn-by-turn data the Fast Fourier Transform (FFT) is computed as demonstrated in Fig.~\ref{fig:hl_FFT}. In the former case (Fig.~\ref{subfig:fft_600}), apart from the sidebands (red) around the betatron tune (black), a dipolar excitation is also visible (purple). As the excitation frequency is not in the vicinity of the betatron tune spread, the dipolar effect has no impact on the tune diffusion. In the second case (Fig.~\ref{subfig:fft_3400}), the frequency of one of the sidebands (\(\rm \approx \)6.89~kHz) exceeds the Nyquist frequency of the turn-by-turn acquisitions (\(\rm \approx\)5.62~kHz) and it is aliased into the spectrum (\(\rm \approx\)4.36~kHz). The dipolar excitation (purple) approaches the betatron tune spread and eventually, has an impact on the betatron motion.

\begin{figure}
\includegraphics[width = \columnwidth]{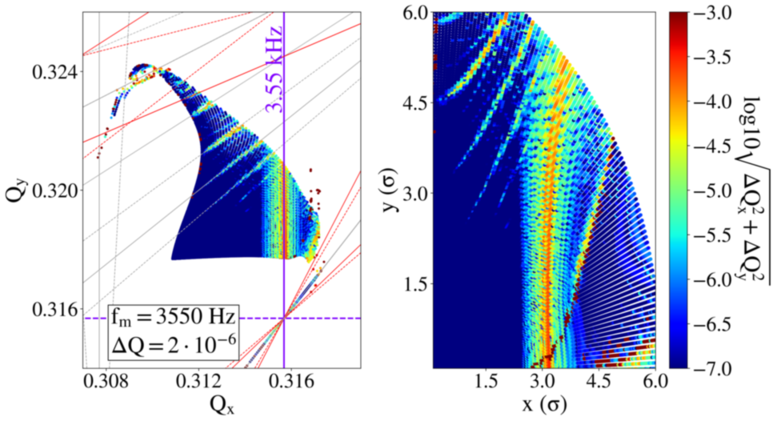} 
\caption{\label{fig:hl_fma_3550} The 5D FMA for a tune modulation at 3.5~kHz. The purple line represents the first-order sideband of the first-order resonance, which is the dipolar excitation at 3.5~kHz in the horizontal plane.}
\end{figure}

The dipolar effect of the triplet eventually leads to the excitation of additional resonances for modulation frequencies close to the tune. These resonances appear in constant frequencies rather than as sidebands around the nominal. This conclusion can also be derived from Eq.~\eqref{eq:resonances} for $m=1, \ l=0$ and $k=1$, which is the excitation due to the dipolar effect. Figure~\ref{fig:hl_fma_3550} illustrates the 5D FMA for a tune modulation at a frequency in the vicinity of the working point (3.5~kHz). In this case, the main contributor to the increase in tune diffusion is the first sideband of the first-order resonance and thus, the dipolar excitation.

\subsubsection{A simple tool to predict frequency sensitivity}

The parametric scan in the tune modulation parameter space presented in Fig.~\ref{fig:hl_heat} is computationally challenging as it requires performing a large number of simulations. Moreover, the tracking must be repeated if the selected working point is modified. Although running the simulations up to $\rm 10^6$ turns is essential to determine the minimum modulation depth that leads to a DA reduction, it is not required to identify, in a fast way, the safest modulation frequencies for operation. In this context, the aim of this section is to present a simple tool to predict the beam sensitivity to specific modulation frequencies based on the position of the sideband resonances. In fact, only the knowledge of the betatron tune spread is needed for the unperturbed case, i.e., the reference conditions without power supply ripples.

\begin{figure}
\subfloat{\subfigimg[width=0.98\columnwidth]{\textbf{a)}}{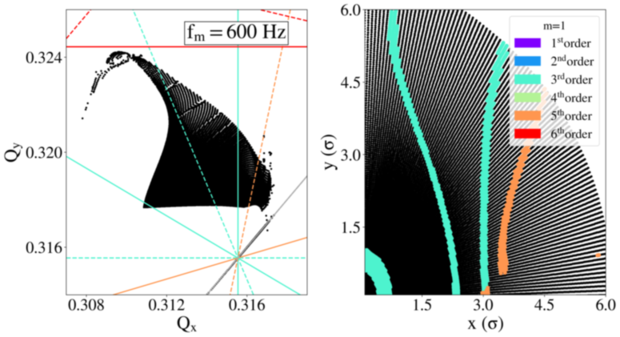} \label{subfig:simple_tool_600Hz}} \\
\subfloat{\subfigimg[width=0.98\columnwidth]{\textbf{b)}}{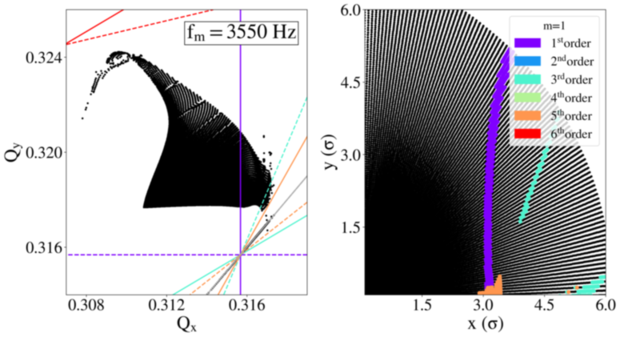} \label{subfig:simple_tool_3550Hz}} 
\caption{\label{fig:simple_tool} The footprint (left panel) and the initial configuration space (right panel) in the absence of power supply ripples. The first-order sideband ($m=1$) of the first (purple), second (blue), third (cyan), fourth (green), fifth (orange) and sixth (red) order resonance ($|k| + |l| $) is illustrated for a modulation at (a) 600~Hz and (b) 3.55~kHz.}
\end{figure}

The working principle of the method is presented in Fig.~\ref{fig:simple_tool}, which shows the betatron tune spread (left panel) and the initial configuration space (right panel) in the absence of power supply ripples (black). Based on the modulation tune, the first-order sideband resonances ($m=1$) from the first to the sixth resonance order ($|k| + |l| \leq 6 $) are computed from Eq.~\eqref{eq:resonances}. Similarly to the previous frequency maps, a different color code is assigned to each resonance order. For instance, Fig.~\ref{subfig:simple_tool_600Hz} and Fig.~\ref{subfig:simple_tool_3550Hz} depict the sideband resonances in the vicinity of the footprint due to a modulation at 600~Hz and 3.55~kHz, respectively. For the selected working point, the first sideband of the third-order resonance (cyan) reaches the footprint for modulation at 600~Hz (Fig.~\ref{subfig:simple_tool_600Hz}). The review of Fig.~\ref{subfig:simple_tool_3550Hz} shows that the first-order sideband of the first-order resonance (purple), which is the dipolar excitation at 3.55~kHz, will affect the footprint. As lower-order resonances have a more significant impact, a tune modulation at 3.55~kHz (first sideband of first-order resonance) is expected to be more detrimental to the beam performance than the modulation at 600~Hz (first sideband of third-order resonance). Then, depending on the amplitudes of the particles that are affected by the resonances in tune domain, a resonance map is performed from the tune domain to the configuration space (right panel). The representation of the resonances in configuration space allows us to more easily determine whether the resonance will affect the tails or the core of the distribution. For instance, a modulation at 3.55~kHz will mainly affect the tails of the distribution, while a modulation frequency of 600~Hz (Fig.~\ref{subfig:simple_tool_600Hz}) will also have an impact on the core of the distribution. A comparison of the predictions of Fig.~\ref{subfig:simple_tool_600Hz} to Fig.~\ref{subfig:fma_600_1}, which is computed with tracking simulations, shows that there is a good agreement between the two. Similarly, a good agreement is found between Fig.~\ref{subfig:simple_tool_3550Hz} and Fig.~\ref{fig:hl_fma_3550}. Therefore, using Eq.~\eqref{eq:resonances}, determining whether the first sideband of the resonances up to a specific order will reach the footprint and mapping these sideband resonances in configuration space allows identifying the ones that will affect the beam depending on the modulation frequency. 

\begin{figure}
\includegraphics[width = \columnwidth]{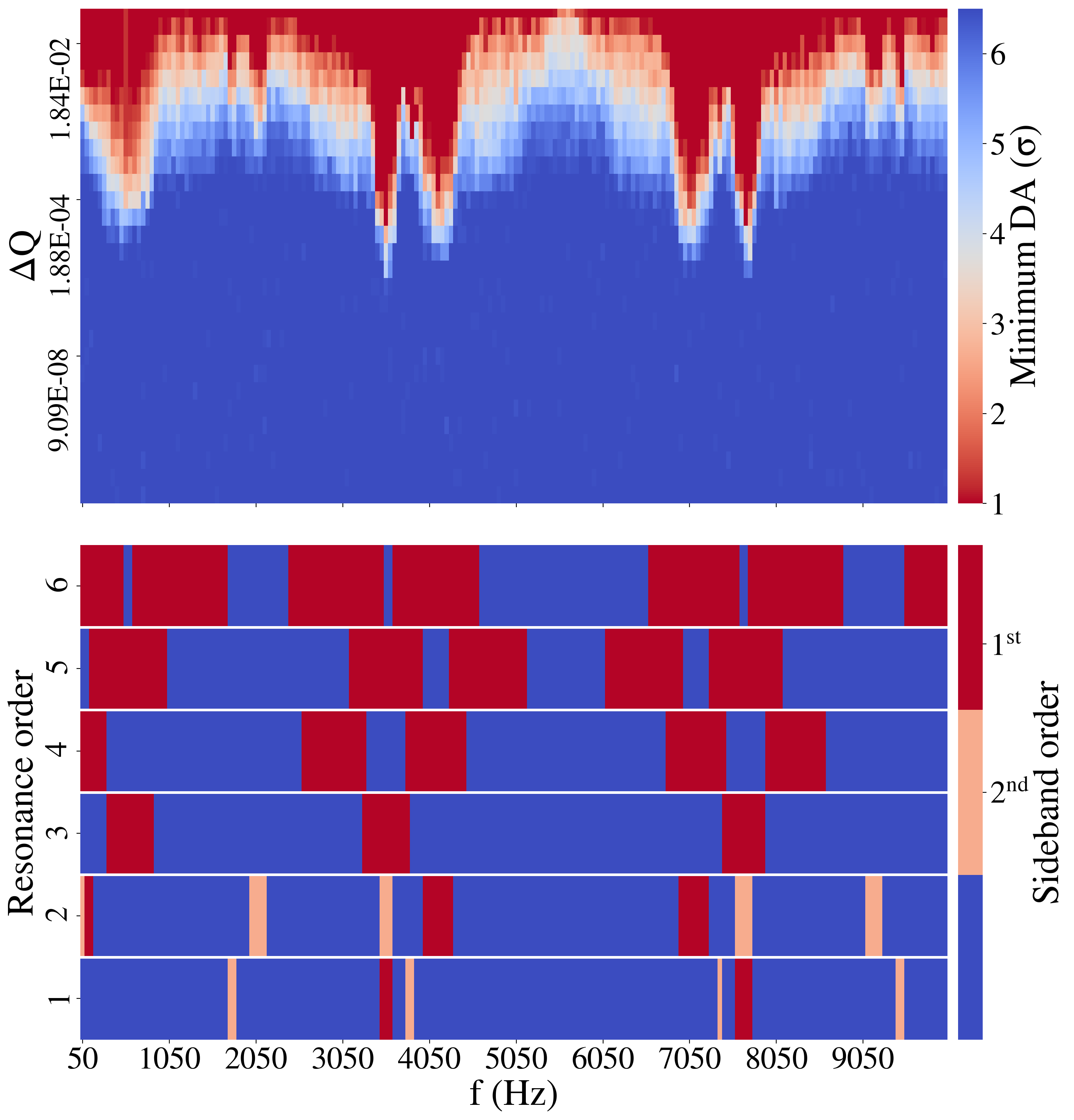} 
\caption{\label{fig:simple_tool_binary} The tune modulation parameter space color-coded with the minimum DA (top) and the modulation frequency as a function of the resonance order (bottom). A color-code is assigned (bottom) depending on whether the first (red) or second (orange) order sideband of the resonance order reaches the footprint or not (blue) for each modulation frequency.}
\end{figure}

As a next step, a similar analysis to Fig.~\ref{fig:simple_tool} is performed for all the multiples of 50~Hz up to 10~kHz. Figure~\ref{fig:simple_tool_binary} presents the modulation frequency as a function of the resonance order ($|k| + |l|$) (bottom), while the heat-map of Fig.~\ref{fig:hl_heat} (top) is also included for an easier comparison between the two. For each modulation frequency, the previously described procedure is repeated to determine whether the first or second-order sideband up to the sixth or second resonance order, respectively, is located in the vicinity of the working point. Then, a color-code (bottom) is assigned depending on whether the first (red) or second (orange) order sideband will affect the distribution. The frequency regimes with a red color, especially for low order resonances, are the modulation frequencies with the largest impact on the beam and should, therefore, be avoided. The orange regimes illustrate the second-order sidebands and should be avoided only for large modulation depths. A blue color is assigned to the study if none of the resonances under consideration reach the footprint.

A comparison between the predictions (bottom) and the results of the tracking simulations (top) shows a good agreement between the two. In particular, it is evident that the large impact on DA close to the betatron tune and its alias is due to the first sideband of the first-order resonance. In the vicinity of the betatron tune ($\rm \approx$4.05~kHz) and its alias ($\rm \approx$7.05~kHz), there is an additional high-sensitivity regime due to the first sideband of the second-order resonance. In addition, the impact on DA for modulation frequencies around 50-1050~Hz is attributed to the first sidebands of the second and third-order resonance that reach the footprint for such modulation frequencies. Furthermore, for large modulation depths, a DA reduction is observed in the frequency region close to ($\rm \approx$2.05~kHz), which is due to the second-order sideband of the first and second-order resonance. Finally, no impact on the DA was observed from the tracking simulations around the regime of 6.05~kHz, which is explained by the fact that no sideband up to the sixth order affects the footprint for such modulation frequencies.

In this way, the modulation frequencies with sideband resonances that do not reach the footprint can be easily identified and they can be distinguished in a fast way from the modulation frequencies with a critical impact. This simple tool can guide the selection of the modulation frequencies such as the power supply switching frequencies for several working points. Although this method takes into account the position of the sidebands and the order of the resonance that affects the footprint due to the modulation, it does not consider the resonance overlap between the nominal and the sideband resonances. Due to this limitation, the simulations presented in the previous sections must be performed to identify the modulation frequencies with the most critical impact on the beam performance. In addition, this method is applicable for relatively small modulation depths as only the first and second-order sidebands of the modulation are considered but it can also be extended for a more aggressive power supply noise scenario.

\subsection{Including a power supply noise spectrum with multiple tones}

As demonstrated from studies conducted in the past \cite{HERA3}, the existence of multiple voltage tones in the power supply noise spectrum is more critical than considering individual frequencies with an equivalent modulation depth due to the resonance overlap. To this end, the combined effect of the several voltage tones that are anticipated in the power supply noise spectrum is computed. As mentioned in the previous sections, the envisaged power supply switching frequencies lie in a high-frequency regime and, as they are not expected to perturb the beam motion \cite{Gamba_2019, Martino}, they are not considered in the following analysis. In this context, the power supply noise spectrum under consideration consists of the low-order 50~Hz harmonics. Table~\ref{tab:simulation_parameters_depth} shows the amplitude of each frequency as defined from the root mean square and the maximum modulation depth shown in Fig.~\ref{fig:hl_heat} (dashed lines).

\begin{figure}
\subfloat{\subfigimg[width=0.98\columnwidth]{\textbf{a)}}{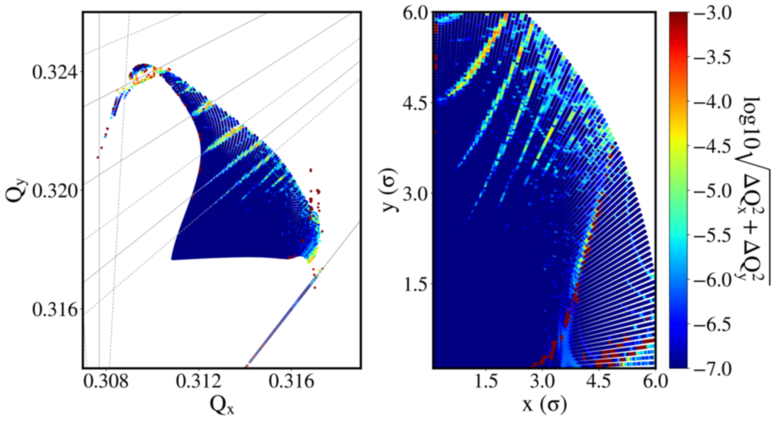} \label{subfig:hllhc_fma_nonoise}} \\
\subfloat{\subfigimg[width=0.98\columnwidth]{\textbf{b)}}{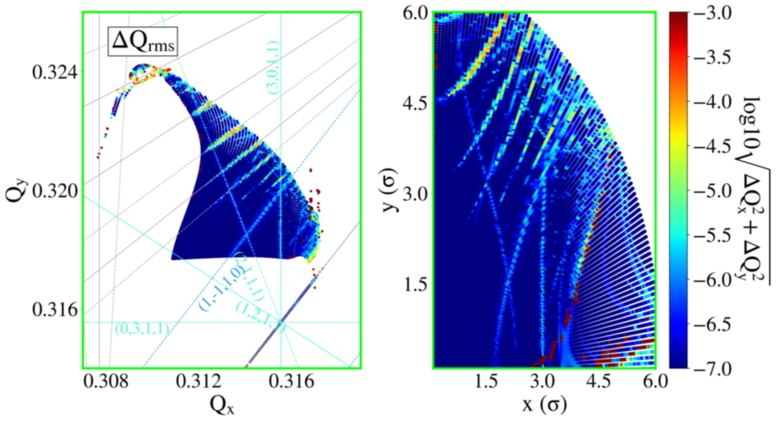} \label{subfig:hllhc_fma_rms}} \\
\subfloat{\subfigimg[width=0.98\columnwidth]{\textbf{c)}}{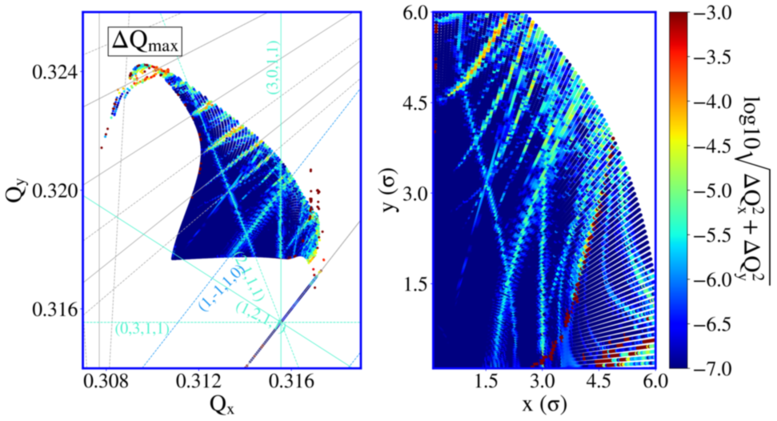} \label{subfig:hllhc_fma_max}} 
\caption{\label{fig:hl_fma_upto600_rms_max} The 5D FMAs (left panel) and the initial configuration space (right panel) (a) in the absence of power supply ripples, including a tune modulation with spectral components at 50, 150, 300 and 600~Hz with (b) the root mean square and (c) the maximum modulation depth as computed from the power supply specifications. The first-order sidebands due to the combined effect of the tones at 50~Hz (blue lines, second-order resonance) and 600~Hz (cyan lines, third-order resonance) are shown.}
\end{figure}

Figure~\ref{fig:hl_fma_upto600_rms_max} depicts the 5D FMAs for three studies: first, with the reference conditions, i.e., in the absence of power supply ripples (Fig.~\ref{subfig:hllhc_fma_nonoise}), second, with amplitudes that correspond to the root mean square modulation depth (Fig.~\ref{subfig:hllhc_fma_rms}) and last, using the maximum modulation depth (Fig.~\ref{subfig:hllhc_fma_max}) for each frequency as computed from the power supply specifications. From the review of the frequency maps the combined impact of the modulation at 50~Hz (blue, first sideband of the second-order resonance) and 600~Hz (cyan, first sideband of the third-order resonance) is visible. However, the increase in tune diffusion is small compared to the reference case. 

\begin{table}
\caption{
The modulation depths of the low order 50~Hz harmonics included in the simulations, as computed from the maximum output voltage of the power supply specifications.
}
\label{tab:simulation_parameters_depth}
\begin{ruledtabular}
\begin{tabular}{ccc}
\textrm{Frequency (Hz)}& \(\Delta Q_{\rm rms} (\times 10^{-6})\) &  \(\Delta Q_{\rm max} (\times 10^{-6})\) \\
\colrule
50 &  \(\rm 1.3 \) & \(\rm 4.5 \)\\
150 &  \(\rm 0.8 \) & \(\rm 2.8\)\\
300 &  \(\rm 0.7 \) & \(\rm 2.4 \)\\
600 &  \(\rm 0.3\) & \(\rm 1.2 \)\\
\end{tabular}
\end{ruledtabular}
\end{table}

\begin{figure}
\includegraphics[width = \columnwidth]{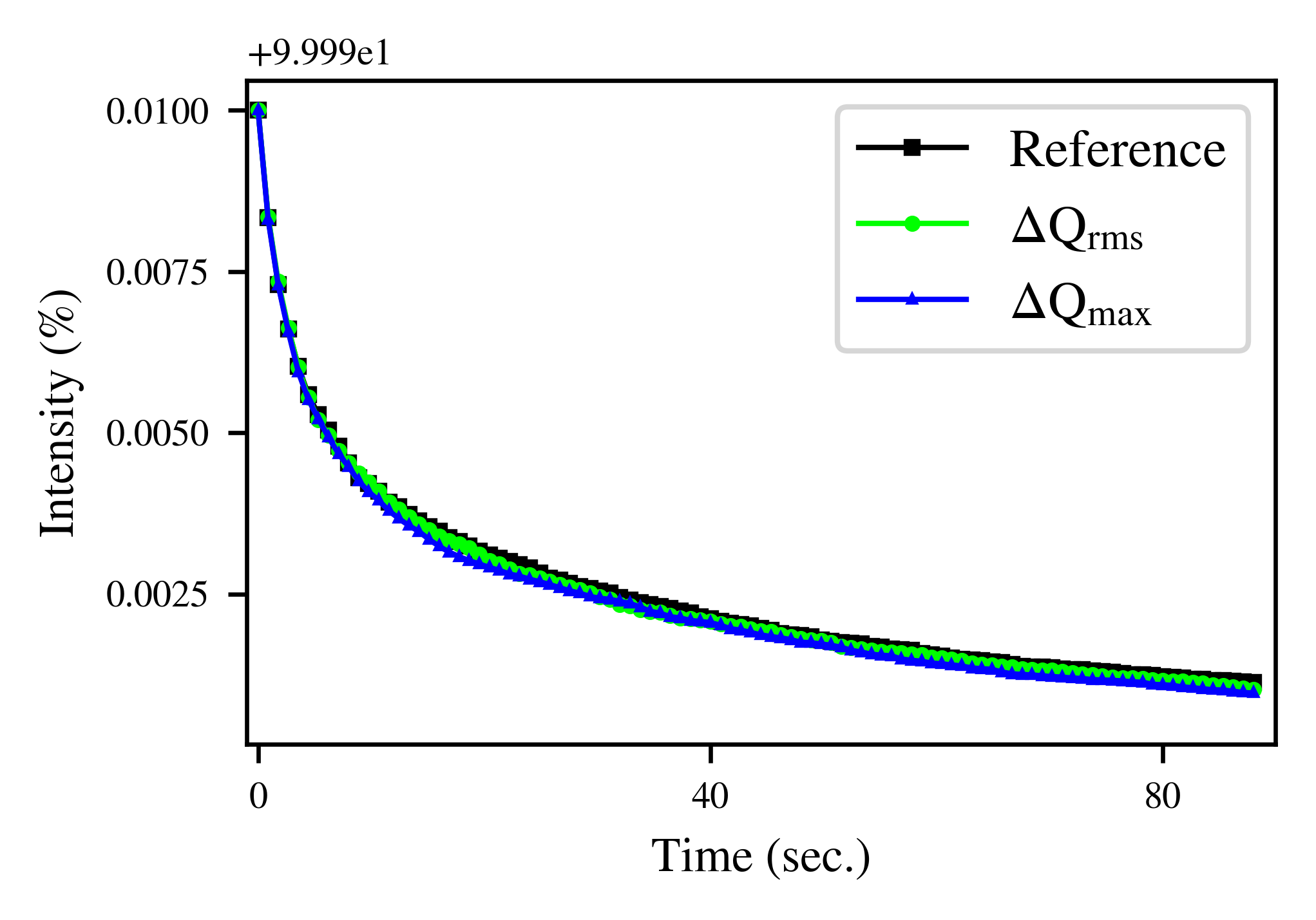} 
\caption{\label{fig:hl_lifetime_upto600_rms_max} Intensity evolution in the absence of power supply ripples (black) and including a power supply noise spectrum that consists of 50, 150, 300 and 600~Hz with the root mean square (green) and maximum (blue) modulation depth.}
\end{figure}

Additional simulations are performed to estimate whether the slight increase in tune diffusion observed in the frequency maps can eventually enhance the particle losses. Figure~\ref{fig:hl_lifetime_upto600_rms_max} illustrates the intensity evolution for the three aforementioned cases. The results suggest that the considered power supply noise spectrum (green and blue) has an insignificant impact on the intensity compared to the reference case (black) for tracking simulations that correspond to 90~seconds of operation. Based on these results, it is concluded that the combined effect of multiple voltage tones in the power supply noise spectrum, with modulation depths that are extracted from the power supply specifications, will not affect the beam performance of the HL-LHC. 

The tracking simulation studies presented in this paper include the most important non-linear fields in the machine such as the ones induced by beam-beam effects and non-linear magnets. Additional effects such as electron-cloud and magnet imperfections, which have been experimentally observed in the LHC, can potentially introduce additional non-linearities \cite{ecloud}. In this context, the interplay of such mechanisms with the power supply ripples should be further investigated in the future.

\section{Conclusions}

This paper investigated the impact of power supply ripples on the beam performance in the HL-LHC inner triplet circuits of IP1 and 5. The increase of the $\beta$-functions in the inner triplets, resulting from the decrease of the beam size at the two IPs and the presence of strong non-linearities, combined with the new hardware that is currently being developed, justifies the need to perform a complete analysis on the repercussions of such effects in future operation. In particular, the predictions shown in this paper are based on simulations with parameters that correspond to the end of the luminosity levelling, a scenario that is the most critical in terms of power supply noise.

By individually scanning several modulation frequencies and depths, in the presence of strong non-linear fields such as beam-beam effects, arc sextupoles and octupoles, the impact of the power supply ripples on the DA was computed. The modulation from the ripples was combined with the one induced by the coupling of the longitudinal and transverse plane for high values of chromaticity. A comparison with the power supply specifications, without considering the shielding effects of the magnets that lead to a further attenuation of the ripples, yielded that the maximum expected power supply noise level is several orders of magnitude lower than the DA reduction threshold. 

It was shown with FMAs that the sensitivity of the beam performance to particular modulation frequencies stems from the excitation of sideband resonances that affect the particles' tune diffusion, a mechanism that strongly depends on the working point and the lattice non-linearities. Contrary to dipolar excitations, quadrupolar power supply ripples can impact the beam performance even if the modulation frequency is not in the vicinity of the betatron frequency. In fact, it introduces a frequency-dependent mechanism of tune diffusion, an effect that was illustrated with frequency maps. Although the concept of sideband resonances in the tune diagram has been introduced in the past, it is the first time that the excitation of additional resonances due to modulation effects has been demonstrated with FMAs. 

Based on the position of the sideband resonances, a methodology to determine the safest modulation frequencies for operation in terms of beam performance has been derived. This simple tool provides fast estimations, which can be easily repeated for different beam and machine configurations, and can guide the selection of the modulation frequencies such as the power supply switching frequencies in the inner triplet.

Including in the simulations the simultaneous effect of several voltage tones that are anticipated in the power supply spectrum, with modulation depths computed from the power supply specifications, it was concluded that the impact on the tune diffusion and eventually, on the lifetime is expected to be negligible.

Therefore, based on the results of the simulations, power supply noise in the inner triplet is not expected to act as a mechanism of luminosity degradation in the HL-LHC era. Although the tracking simulations include the most important non-linear fields in the machine, they do not account for all the effects that have been observed experimentally (e.g. electron-cloud) during the LHC operation. These effects can introduce additional non-linearities and their interplay with the power supply noise may cause a further DA reduction. To this end, as a next step, the DA tolerances defined by the simulations must be experimentally verified in future operation through controlled quadrupolar excitations. Finally, to improve the validity of the simulations, a realistic power supply noise spectrum must be included, in combination with a more accurate representation of the transfer function of the power supply noise spectrum from the voltage to the magnetic field.

In the context of this study, a general analysis framework of simulation data has been developed. In particular, this paper demonstrated the importance of FMA in the investigation of noise effects, illustrated methods to define power supply noise thresholds through DA scans and presented tools to determine the intensity evolution with weighted distributions. The aforementioned methods can be applied to studies of several other types of noise effects.

\begin{acknowledgments}
The authors gratefully acknowledge G.~Arduini, R.~De Maria, M. Fitterer, D.~Gamba and M.~Martino for valuable suggestions and discussions on this work. 
\end{acknowledgments}

\FloatBarrier
\bibliography{bibliography}
\end{document}